\documentstyle[preprint,aps,epsfig,eqsecnum]{revtex}
\tightenlines
\def\m@thcombine#1#2{%
  \setbox0=\hbox{$#1$}
  \setbox1=\hbox{$#2$}
  \ifdim\wd0>\wd1
    \setbox0=\hbox to\wd1{\hss\box0\hss}
  \else
    \setbox1=\hbox to\wd0{\hss\box1\hss}
  \fi
  \mathop{\vcenter{
    \offinterlineskip\box0\box1}}}
\def\lesim{\m@thcombine<\sim}
\def\gesim{\m@thcombine>\sim}
\begin{document}
 
\newdimen\picraise
\newcommand\picbox[1]
{
  \setbox0=\hbox{\input{#1}}
  \picraise=-0.5\ht0
  \advance\picraise by 0.5\dp0
  \advance\picraise by 3pt      
  \hbox{\raise\picraise \box0}
}

\draft
\title{ A TWO-DIMENSIONAL QCD IN THE COVARIANT GAUGE}

\author{V. Gogokhia and Gy. Kluge}

\address{HAS, CRIP, RMKI, Depart. Theor. Phys., Budapest 114, P.O.B. 49,
H-1525, Hungary \\
 email addresses: gogohia@rmki.kfki.hu and kluge@rmki.kfki.hu}

\maketitle

\begin{abstract}
A nonperturbative approach to 2D covariant gauge QCD is presented in the
context of the Schwinger-Dyson equations for quark and ghost
propagators and the corresponding Slavnov-Taylor identities.
 The distribution theory, complemented by the dimensional
regularization method, is used in order to correctly treat the
severe infrared singularities which inevitably appear in the
theory. By working out the multiplicative renormalization program
we remove them from the theory on a general ground and in a
self-consistent way, proving thus the infrared multiplicative
renormalizability of 2D QCD within our approach. This makes it
possible to sum up the infinite series of the corresponding planar
skeleton diagrams in order to derive a closed set of equations for
the infrared renormalized quark propagator. We have shown that
complications due to ghost degrees of freedom can be considerable
within our approach. It is shown exactly that 2D covariant gauge
QCD implies quark confinement (the quark propagator has no poles,
indeed) as well as dynamical breakdown of chiral symmetry (a
chiral symmetry preserving solution is forbidden). We also show
explicitly how to formulate the bound-state problem and the
Schwinger-Dyson equations for the gluon propagator and the triple
gauge proper vertex, all free of the severe IR singularities.
\end{abstract}


\vfill

\eject

\section{Introduction}

In his paper [1] 't Hooft investigated two-dimensional (2D) QCD in the
light-cone gauge which is free from ghost complications. He used also
large $N_c$ (the number of
colors) limit technique in order to make the perturbation (PT)
expansion
with respect to $1/N_c$ reasonable. In this case the planar diagrams are
reduced to quark self-energy and ladder diagrams which can be
summed. The bound-state problem within the Bethe-Salpeter (BS) formalism was
finally obtained free from the infrared (IR) singularities. The existence of
a discrete
spectrum only (no continuum in the spectrum) was demonstrated. Since this
pioneering paper 2D QCD continues to attract attention (see, for example,
review
[2] and recent papers [3-5] and references therein). Despite its simplistic
 vacuum structure it remains a rather good laboratory for the modern
theory of strong interaction which is four-dimensional (4D) QCD [6].

The most important yet unsolved problems in QCD are, of course,
quark confinement and the dynamical
(spontaneous) breakdown of chiral symmetry (or equivalently
dynamical chiral symmetry breaking (DCSB)) closely related to it.
In this work a new,
nonperturbative (NP) solution (using neither large $N_c$ limit
technique explicitly nor a weak coupling regime, i.e., ladder
approximation) to 2D QCD in the covariant gauge is
obtained. This makes it possible to construct a 2D
covariant gauge model for the above-mentioned important phenomena.
It is well known, however, that covariant gauges, in general, are
complicated by the ghost contributions. Nevertheless, we will show
that ghost degrees of freedom can be considerable within our
approach. The ghost-quark sector contains a very important piece of
information on quark degrees of freedom themselves through the
corresponding quark Slavnov-Taylor (ST) identity. This is just the
information which should be self-consistently taken into account.
Some results of the present investigation have been already presented
in Ref. [7].

The paper is organized as follows. In section II we derive the IR
renormalized Schwinger-Dyson (SD) equation for the quark
propagator. In section III the SD equation for the IR renormalized
ghost self-energy is also derived. In section IV the quark-ghost
sector represented by the quark ST identity is analyzed and the IR
renormalized quark ST identity is obtained. In section V we show
that the obtained complete set of equations for the IR
renormalized quark propagator can be reduced to a system of
coupled, nonlinear differential equations of the first order. By
solving the above-mentioned system of equations, it is explicitly
shown that the quark propagator has no poles, indeed (section VI),
and that the dynamical (spontaneous) breakdown of chiral symmetry
is required (section VII). In section VIII the IR properties of
the theory in the quark-ghost and Yang-Mills (YM) sectors (by
using  the corresponding ST identities for the three- and
four-gluon vertices) has been discussed. Within the BS formalism
we formulate the bound-state problem free from the IR
singularities. In sections IX and X the IR properties of the SD
equations for the gluon propagator and three-gluon proper vertex
are investigated, respectively. This makes it possible to
formulate a general system in order to remove all the severe IR
divergences from the theory in a self-consistent way,
 and thus to prove the IR multiplicative renormalizability of
our approach to 2D QCD. In section XI we compare our approach with
the 't Hooft model [1] with respect to the approximations made. In
section XII we discuss our results and present our conclusions.
Some perspectives for 4D QCD are also discussed there. We have
investigated a nonzero quark masses case in appendix A. We have
shown explicitly that our solution for the quark propagator
possesses a heavy quark flavor symmetry. In appendix B we show
schematically how the bound-state problem can be reduced to an
algebraic problem within our approach in the framework of the BS
formalism.

\section{IR renormalized quark propagator}

   Let us consider the SD equation for the quark propagator (PT unrenormalized
(as well as other quantities) for simplicity in order not
to complicate notations here and everywhere below) in momentum space with
Euclidean signature (see Fig. 1)

\begin{equation}
S^{-1}(p) = S^{-1}_0(p) - g^2 C_F i \int {d^nq\over {(2\pi)^n}}
\Gamma_\mu(p, q) S(p-q)\gamma_\nu D^0_{\mu\nu}(q),
\end{equation}
where  $C_F$  is the eigenvalue of the quadratic Casimir operator
in the fundamental representation (for $SU(N_c)$, in general, $C_F
= (N^2_c - 1)/2N_c = 4/3$) and

\begin{equation}
S^{-1}_0(p) = i (\hat p + m_0)
\end{equation}
with $m_0$ being the current ("bare") mass of a single quark.
$\Gamma_\mu(p,q)$ is the corresponding quark-gluon proper vertex
function. Instead of the simplifications due to the limit $N_c
\rightarrow \infty$ at fixed $g^2 N_c$ and light-cone gauge [1,8]
(see section XI below), we are going to use throughout the present
investigation the free gluon propagator in the covariant gauge
from the very beginning. This makes it possible to maintain the
direct interaction of massless gluons which is the main dynamical
effect in QCD of any dimensions.
 In the covariant gauge it is

\begin{equation}
D^0_{\mu\nu}(q) =i \Bigl( g_{\mu\nu} +(\xi -1){q_\mu q_\nu \over
q^2} \Bigr)
 {1 \over q^2},
\end{equation}
where $\xi$ is the gauge fixing parameter. Let us emphasize the
fact that by using the gluon propagator in the whole momentum
range, we are investigating the quark propagator in the whole
momentum range as well.

The important observation now is that for the free gluon
propagator the exact singularity $1/q^2$ at $q^2 \rightarrow 0$ in
2D QCD is severe and therefore it should be correctly treated
within the distribution theory (DT) [9,10] (in Ref. [10] some
fundamental results of pure mathematical tractate on the DT [9]
necessary for further purpose are presented in a suitable form).
In order to actually define the system of the SD equations (see
below) in the IR region, it is convenient to apply the
gauge-invariant dimensional regularization (DR) method of 't Hooft
and Veltman [11] in the limit $D=2 + 2 \epsilon, \ \epsilon
\rightarrow 0^+$. Here and below $\epsilon$ is the IR
regularization parameter which is to be set to zero at the end of
computations. Let us use in the sense of the DT (i.e., under
integrals, taking into account the smoothness properties of the
corresponding test functions) the relation [9,10]

\begin{equation}
 q^{-2} = { \pi \over \epsilon} \delta^2(q) + finite \ terms,
\quad \epsilon \rightarrow 0^+.
\end{equation}
We point out that after introducing this expansion here and
everywhere below, one can fix the number of dimensions, i.e., put
$D=n=2$ without any further problems since there will be no other
severe IR singularities with respect to $\epsilon$ as $\epsilon
\rightarrow 0^+$ in the corresponding SD equations but those
explicitly shown in this expansion.

\begin{figure}[bp]

\[ \input{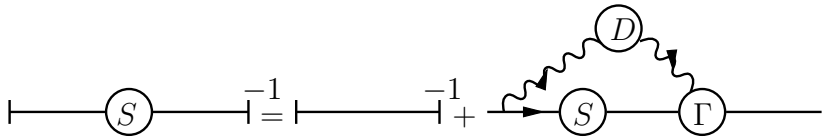}  \]

\caption{The quark SD equation. Here and in all figures below
$D\rightarrow D^0$ is understood. }
\label{autonum}
\end{figure}

It is worth emphasizing that the IR singularity (2.4) is, on the
one hand, a unique, simplest IR singularity possible in 2D QCD; on
the other hand, it is a NP (severe) singularity at the same time
[9,10]. In this connection, let us remind that in 4D QCD the free
gluon's IR singularity is not severe, i.e., the Laurent expansion
(2.4) does not exist in this case, so it is a PT singularity
there. In other words, the free gluon propagator is the NP itself
from the very beginning, and thus may serve as a rather good
approximation to the full gluon propagator, at least in the deep
IR region, since it exactly reproduces a possible severe IR
singularity of the full gluon propagator. This is important since
precisely the IR properties of the theory are closely related to
its NP dynamics, and therefore they are responsible for such NP
effects as quark confinement and dynamical (spontaneous) breakdown
of chiral symmetry. That the free gluon propagator IR singularity
exactly reproduces a possible simplest NP IR singularity of the
full gluon propagator, is a particular feature of 2D QCD. This
underlines a special status of this theory. In this case all other
Green's functions (in particular, the quark-gluon and ghost-gluon
vertices) should be considered as regular functions of the
momentum transfer (otherwise, obviously, the IR singularity
becomes effectively stronger than (2.4)). In the quark-ghost
sector, however, the momentum transfer goes through the momentum
of the ghost self-energy (see section IV). In its turn, this means
that the quark-gluon vertex is regular with respect to the ghost
self-energy momentum. At the same time, we will show that the
ghost self-energy can be regular at the origin as well.
Apparently, in nD QCD all the severe IR singularities are to be
mainly accumulated in the full gluon propagator and effectively
correctly described by its structure in the IR domain.

In the presence of such a severe singularity (2.4) all Green's
functions become generally dependent on the IR regularization
parameter $\epsilon$, i.e., they become IR regularized. For
simplicity, this dependence is not shown explicitly. Let us
introduce the IR renormalized quark-gluon vertex function,
coupling constant and the quark propagator as follows:

\begin{eqnarray}
\Gamma_\mu(p, q) &=& Z^{-1}_1(\epsilon) \bar \Gamma_\mu(p, q), \nonumber\\
g^2 &=& X(\epsilon) \bar g^2, \qquad  \epsilon \rightarrow  0^+, \nonumber\\
S(p) &=& Z_2( \epsilon) \bar S(p).
\end{eqnarray}
Here and below $Z_1(\epsilon), \ Z_2(\epsilon)$ and $X(\epsilon)$
are the corresponding IR multiplicative renormalization (IRMR) constants.
The $\epsilon$-parameter dependence is indicated explicitly
to distinguish them from the usual ultraviolet (UV) renormalization
constants. In all relations containing the IRMR constants, the
$ \epsilon \rightarrow  0^+ $ limit is always assumed at the final stage.
$\bar \Gamma_\mu(p, q)$ and $\bar S(p)$ are the IR renormalized Green's
functions
and therefore they do not depend on $\epsilon$ in the
$ \epsilon \rightarrow  0^+ $ limit, i.e. they exist as
$ \epsilon \rightarrow  0^+ $, as does the IR renormalized coupling
constant
$\bar g^2$ (charge IR renormalization). There are no restrictions on the
$ \epsilon \rightarrow  0^+ $ limit behavior of the IRMR
constants apart from the smooth $\epsilon$
dependence of the quark wave function IRMR constant $Z_2(\epsilon)$ (see Eq.
(2.7) below).

Substituting all these relations into the quark SD equation (2.1), and
taking into account the expansion (2.4), we see that a cancellation
of the IR divergences takes place if and only if (iff)

\begin{equation}
X(\epsilon) Z^2_2(\epsilon) Z^{-1}_1(\epsilon)= \epsilon Y_q, \qquad \epsilon \rightarrow 0^+,
\end{equation}
holds. Here $Y_q$ is an arbitrary but finite constant. Thus the relation
(2.6) is the quark SD equation IR convergence condition in the most general
form. It is evident  that  this very condition and the
similar ones below govern the concrete $\epsilon$-dependence of the IRMR
constants which, in general, remain arbitrary.
The quark SD equation for the IR renormalized quantities becomes

\begin{equation}
\bar S^{-1}(p) = Z_2 (\epsilon) S^{-1}_0(p) + \bar{g}^2 Y_q \bar \Gamma_\mu(p, 0)
\bar S(p)\gamma_\mu.
\end{equation}
Let us note that the IR renormalized coupling constant in 2D QCD
has the dimensions of mass. All other finite numerical factors
(apart from $Y_q$) have been included into it. Also here and below
all other finite terms become terms of order $\epsilon$ and
therefore they vanish in the $\epsilon \rightarrow 0^+$ limit
after the completion of the IRMR program (in order to remove all
the severe IR singularities from the theory on a general ground).

A few remarks are in order. Here and everywhere below in the derivation of the
equations for the IR renormalized quantities we use the relation $q_\mu
q_\nu = (1/2) g_{\mu\nu} q^2$ in the sense of the
symmetric integration in 2D Euclidean space since it is multiplied
by the $\delta$ function (i.e., $q \rightarrow 0$). As was mentioned
above,
the finite numerical factor $(\xi +1)/2$ has been included into the IR
renormalized coupling constant (in principle,
in the presence of an arbitrary mass scale parameter one can
forget about arbitrary, finite constants). In its turn,
this means that there is no explicit dependence on the gauge fixing
parameter in the quark SD equation (2.7). The same will be true for the
quark ST identity (see below).

Let us also show briefly that the gauge fixing parameter is the IR finite from
the very beginning, indeed. Similar to relations (2.5), let us introduce the
IRMR
constant of the gauge fixing parameter as follows: $\xi = X_1(\epsilon) \bar \xi$,
where again $\bar \xi$ exists as $\epsilon$ goes to zero, by definition.
Then in addition to the quark SD equation IR convergence condition (2.6) one
has
one more condition including the gauge fixing parameter IRMR constant, namely

\begin{equation}
X_1(\epsilon) X(\epsilon) Z^2_2(\epsilon) Z^{-1}_1(\epsilon)= \epsilon Y_1,
\qquad \epsilon \rightarrow 0^+.
\end{equation}
However, combining these two conditions, one immediately obtains $X_1(\epsilon)
=X_1 = Y_1 Y_q^{-1}$. So this finite but arbitrary number can be put to
unity not
losing generality since nothing depends explicitly on the gauge fixing
parameter.

 The information about the quark-gluon vertex function at zero momentum
transfer can be provided by the quark ST identity [6,12,13] which
contains
unknown ghost contributions in the covariant gauge. For this reason let us
consider in the next section the SD equation for the ghost self-energy.

\section{IR renormalized ghost self-energy}

     The ghost self-energy $b(k^2)$   also obeys a simple
SD equation in Euclidean space [6,14] (see Fig. 2)

\begin{equation}
i k^2 b( k^2) = g^2 C_A i \int {d^nq\over {(2\pi)^n}}
G_\mu(k,q) G(k-q) (k-q)_\nu D^0_{\mu\nu}(q),
\end{equation}
where $C_A$ is the eigenvalue of the quadratic Casimir
operator in the adjoint representation (for $SU(N_c)$, in general,
 $C_A = N_c$). The ghost propagator is

\begin{equation}
G(k) = - {i  \over k^2 [ 1 + b(k^2)]}
\end{equation}
and

\begin{equation}
G_\mu(k,q) = k^\lambda G_{\lambda\mu}(k,q)
\end{equation}
is the ghost-gluon vertex function ( $G_{\lambda\mu} =
g_{\lambda\mu}$ in  perturbation theory).

\begin{figure}[bp]

\[ \input{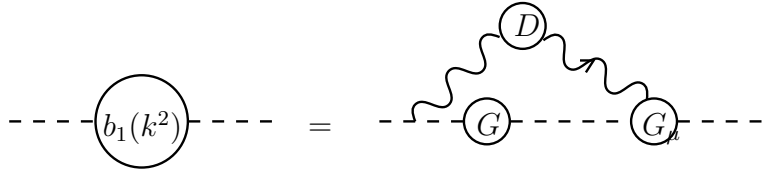}  \]

\caption{The ghost self-energy SD equation with definition
$b_1(k^2)= ik^2 b(k^2)$ .}
\label{autonum}
\end{figure}

   Similar to the previous relations, let us introduce the IR renormalized
ghost self-energy

\begin{equation}
b(k^2) = \tilde{Z}(\epsilon) \bar b(k^2), \qquad \epsilon \rightarrow  0^+
\end{equation}
and the IR renormalized ghost-gluon vertex function

\begin{equation}
G_\mu(k,q) = \tilde{Z}_1(\epsilon) \bar G_\mu(k,q), \qquad \epsilon \rightarrow  0^+,
\end{equation}
where $\bar b(k^2)$ and  $\bar G_\mu(k,q)$ are IR renormalized, by
definition. Thus they do not depend on the parameter $\epsilon$ in the
$\epsilon \rightarrow  0^+$ limit which is always assumed in this kind of
relations. $\tilde{Z}(\epsilon)$ and $\tilde{Z}_1(\epsilon)$ are the
corresponding IRMR constants. The IR renormalized ghost propagator is defined
as

\begin{equation}
G(k) = \tilde{Z}_2(\epsilon) \bar G(k), \qquad \epsilon \rightarrow  0^+,
\end{equation}
where $\tilde{Z}_2(\epsilon)$ is also the corresponding IRMR constant and
$\bar G(k)$ exists as $ \epsilon \rightarrow  0^+$.
From these definitions it follows that the ghost propagator
IRMR constant $\tilde{Z}_2(\epsilon)$ is completely determined by the ghost
self-energy IRMR constant $\tilde{Z}(\epsilon)$ and vice versa, i.e.,

\begin{equation}
\tilde{Z}_2(\epsilon) = \tilde{Z}^{-1}(\epsilon).
\end{equation}
As in the previous case, the dependence of these IRMR constants on $\epsilon$
in general is arbitrary apart from the ghost self-energy IRMR constant $\tilde{Z}(\epsilon)$.
 The expression for the IR renormalized ghost propagator is

\begin{equation}
\bar G(k) = - {i \over {k^2\left[\tilde{Z}^{-1} (\epsilon) + \bar b (k^2) \right]}},
\qquad  \epsilon \rightarrow  0^+.
\end{equation}
From this expression it obviously follows that the regular dependence of
$\tilde{Z}(\epsilon)$ on $\epsilon$ in the $\epsilon \rightarrow  0^+$ limit
should be excluded from the very beginning. The
problem is that if $\tilde{Z}(\epsilon)$ vanishes as
$ \epsilon \rightarrow  0^+$, i.e., $\tilde{Z}^{-1}(\epsilon)$ is singular,
then the full ghost propagator simply reduces to the free one, and there is no
nontrivial renormalization at all.
In other words, in this case the IR renormalized ghost propagator is vanishing
in the $ \epsilon \rightarrow  0^+$ limit. This means in its turn
that all the
necessary information about quark degrees of freedom which is contained in the
quark-ghost sector will be finally totally lost (see next section). Thus the
only nontrivial cases remaining are:

1). When the ghost
self-energy is IR renormalized from the very beginning
(i.e., $\tilde{Z}(\epsilon) \equiv \tilde{Z} = const.$, so that it is IR
finite), then the ghost propagator is also IR finite.

2). The IRMR constant $\tilde{Z}(\epsilon)$ is singular as $\epsilon$ goes to
zero, so its inverse is regular in the same limit.

Substituting all these relations as well as relation (2.4) into
the initial SD equation for the ghost self-energy (3.1), we see
obtains that a cancellation of the severe IR divergences takes
place iff

\begin{equation}
X(\epsilon) \tilde{Z}_1(\epsilon) \tilde{Z}_2(\epsilon) \tilde{Z}^{-1}(\epsilon) = \epsilon Y_g,
 \qquad  \epsilon \rightarrow  0^+,
\end{equation}
holds. Here $Y_g$ is an arbitrary but finite constant (different
from $Y_q$, of course). This is the ghost self-energy SD equation
IR convergence condition in the most general form. The ghost SD
equation for the IR renormalized quantities becomes (in the Euclidean
space)

\begin{equation}
ik^2 \bar b(k^2) = - \bar g^2_1  Y_g \bar G_\mu(k,0) \bar G(k) k_{\mu},
\end{equation}
where all known finite numerical factors are included into
the IR renormalized coupling constant $\bar g_1^2$, apart from $Y_g$
(see section IV).

\subsection{Ghost-gluon vertex}

 In order to show that the IR renormalized ghost self-energy may exist and be
finite at origin, one has to extract $k^2$ from the right hand side of
Eq. (3.10), and then pass to the limit $k^2=0$. For this aim, let us
consider the IR renormalized
counterpart of the ghost-gluon vertex (3.3) which is shown in Eq.
(3.5)

\begin{equation}
\bar G_\mu(k,q) = k^\lambda \bar G_{\lambda\mu}(k,q).
\end{equation}
Its general decomposition is

\begin{equation}
\bar G_{\lambda\mu}(k,q) = g_{\lambda\mu} G_1 + k_\lambda k_\mu G_2 +
q_\lambda q_\mu G_3 +  k_\lambda q_\mu G_4 + q_\lambda k_\mu G_5,
\end{equation}
and ($l=k-q$)

\begin{equation}
G_i \equiv  G_i (k^2, q^2, l^2), \qquad i = 1, 2, 3, 4, 5.
\end{equation}
Substituting this into the previous vertex (3.11), one obtains

\begin{equation}
\bar G_{\mu}(k,q) = k_\mu \bar G_1 (k, q) + q_\mu \bar G_2 (k, q),
\end{equation}
where

\begin{eqnarray}
\bar G_1(k,q) &=& G_1 + k^2 G_2 + (kq) G_5 =  G_1 + k^2 (G_2 + G_5) - (kl) G_5, \nonumber\\
\bar G_2(k,q) &=& k^2 G_4 + (kq) G_3 = k^2 (G_3 + G_4) - (kl) G_3.
\end{eqnarray}
Thus at zero momentum transfer ($q=0$), one has

\begin{equation}
\bar G_{\mu}(k,0) = k_\mu \bar G_1 (k, 0) = k_\mu \bar G_1 (k^2),
\end{equation}
where

\begin{equation}
\bar G_1(k^2) = G_1 (k^2) + k^2 G_2 (k^2).
\end{equation}
Let us remind that the form factors (3.13) exist when any of their
momenta goes to zero.\footnote{The significance of the unphysical
kinematical singularities in the Euclidean space, where $k^2=0$
implies $k_i=0$, becomes hypothetical. In Minkowski space they
always can be removed in advance by the Ball and Chiu procedure
[15] as well as from the quark-gluon vertex.} Taking now into
account the relation (3.16) and the definition (3.8), it is easy
to see that the corresponding equation (3.10) for determining
$\bar b(k^2)$ is nothing else but an algebraic equation of
second order, namely

\begin{equation}
\bar b^2(k^2) + \tilde{Z}^{-1} \bar b (k^2) = { 1 \over k^2} \bar g_1^2 Y_g
\bar G_1(k^2).
\end{equation}

Its solutions are
\begin{equation}
\bar b_{1,2}(k^2) = - {1 \over 2} \tilde{Z}^{-1} \pm \sqrt{{1 \over 4} \tilde{Z}^{-2} +
{ 1 \over k^2} \bar g_1^2 Y_g \bar G_1(k^2)}.
\end{equation}
Let us remind that in this equation $\tilde{Z}^{-1} \equiv \tilde{Z}^{-1} (\epsilon)$ is
either constant or vanishes as $\epsilon \rightarrow  0^+$,
so it always exists in this limit. If now (see also Ref. [14])

\begin{equation}
\bar G_1 (k^2) = k^2 R_1 (k^2), \quad k^2 \rightarrow 0,
\end{equation}
and $R_1(k^2)$ exists and is finite at zero point, then the
ghost-self energy exists and is finite at the origin as
well.\footnote{In principle, singular dependence of the ghost
self-energy on its momentum should not be excluded $a \ priori$.
However, the ST identity (see next section) is to be treated in a
completely different way in this case and therefore it is left for
consideration elsewhere. Also the smoothness properties of the
corresponding test functions are compromised in this case and the
use of the relation (2.4) becomes problematic, at least in the
standard DT sense.} Because of the relation (3.17) this can be
achieved in general by setting $G_1(k^2)= k^2 R(k^2)$ and then
$R_1(k^2)= R(k^2) + G_2(k^2)$. Let us emphasize in advance that
our final results will not explicitly depend on the auxiliary
technical assumption (3.20).

Obviously, Eq. (3.10) can be rewritten in the equivalent form as follows:

\begin{equation}
- \bar g^2_1 Y_g \bar G_\mu(k,0) \bar G(k) = i k_{\mu} \bar b (k^2),
\end{equation}
then it follows that the right hand side of this
relation is of order $k$ ($\sim O(k)$) always as $k \rightarrow 0$.
Thus the ghost-self energy exists and is finite at zero point but remains
arbitrary within our approach.

Concluding, let us note that, in principle, the information about the
ghost-gluon vertex (3.11) could be obtained from the corresponding
identity derived in Ref. [16]. We found (in complete agreement
with Pagels [14]) that even at zero momentum
transfer no useful information can be obtained, indeed.  It has a too
complicated mathematical
structure and involves the matrix elements of composite operators of ghost
and gluon fields. However, let us emphasize that our approach makes it
possible to avoid this difficulty (see below).

\section{IR renormalized quark ST identity}

    Let us consider the ST identity for the quark-gluon
vertex function $\Gamma_\mu(p,k)$:

\begin{equation}
- i k_\mu \Gamma^a_\mu(p,k) \left[ 1 +
b(k^2) \right] = \left[ T^a - B^a(p,k)\right]
S^{-1}(p+k) -  S^{-1}(p)\left[T^a - B^a(p,k)\right],
\end{equation}
where $b(k^2)$ is the ghost self-energy and $B^a(p,k)$ is the ghost-quark
scattering kernel [6,14,17,18]; $T^a$'s are color group generators.
From it one recovers the standard Ward-Takahashi (WT)
identity in the formal $b = B = 0$  limit.
The ghost-quark scattering kernel $B^a(p,k)$ is determined by its
skeleton expansion

\begin{equation}
B^a (p, k) = \sum_{n=1}^{\infty} B^a_n (p, k)
\end{equation}
which is diagrammatically shown in Fig. 3 (see also Refs. [14,18]).

\begin{figure}[bp]

\[ \input{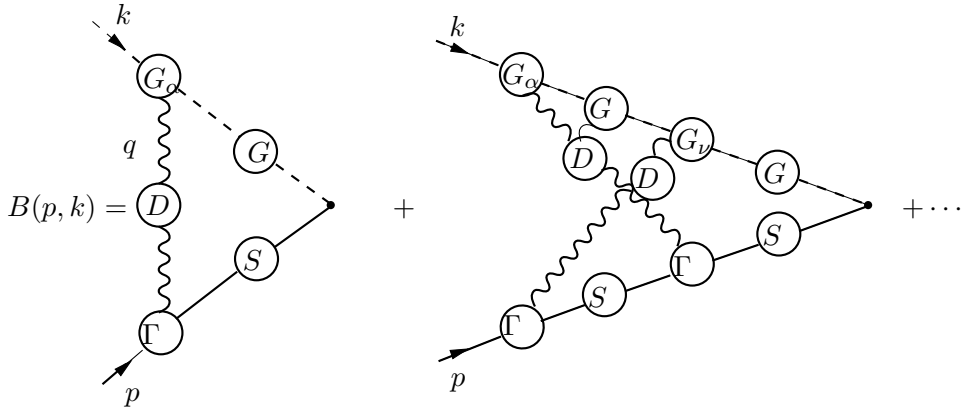}  \]

\caption{The skeleton expansion for the ghost-quark scattering kernel.}
\label{autonum}
\end{figure}

In addition to the previous IR renormalized quantities, it is convenient to
introduce independently the "IRMR constant" for the ghost-quark
scattering
kernel $B^a(p,k)$ itself as follows:

\begin{equation}
B^a(p,k)  = \tilde{Z}_B (\epsilon) \bar B^a (p,k), \qquad \epsilon \rightarrow 0^+.
\end{equation}
Then the IR renormalized version of the quark ST identity (4.1) becomes
(here and below we have already escaped the dependence
on the color group generators $T^a$'s)

\begin{eqnarray}
- i k_\mu \bar \Gamma_\mu(p,k) \left[ \tilde{Z}^{-1}(\epsilon) + \bar b(k^2) \right]
&=& \left[ \tilde{Z}_B^{-1}(\epsilon) - \bar B (p,k) \right] \bar S^{-1}(p+k)
\nonumber\\
&-& \bar S^{-1}(p) \left[ \tilde{Z}_B^{-1}(\epsilon) - \bar B(p,k) \right],
\end{eqnarray}
iff the corresponding quark ST identity IR convergence relation

\begin{equation}
Z_1^{-1}(\epsilon) \tilde{Z}(\epsilon) =  Z_2^{-1}(\epsilon) \tilde{Z}_B (\epsilon),
\qquad \epsilon \rightarrow  0^+,
\end{equation}
holds. Our final results will not depend on the quark-ghost
scattering kernel "IRMR constant". It plays only auxiliary role.
It is almost obvious that this "IRMR constant" does not depend on
$\epsilon$ at all , i.e., $\tilde{Z}_B(\epsilon) =
\tilde{Z}_B=const.$, remaining an arbitrary finite constant.
Otherwise, from the ST identity (4.4) it would simply follow that
either the information about quark degrees of freedom (which is
contained in $\bar B (p,k)$) would be lost (regular dependence) or
the correspondence with the WT identity would be lost (singular
dependence). That is why in what follows we will omit its
dependence on $\epsilon$. Let us note that the IRMR program can be
formulated without explicitly introducing it (see second paper
in Ref. [18]).

Let us start with the investigation of the first term $B_1(p,k)$ in the
$B(p,k)$
skeleton expansion (4.2). After the evaluation of the color group factors
it becomes (Euclidean space)

\begin{equation}
B_1(p,k) = - {1\over 2}
g^2 C_A i \int {d^n q \over {(2\pi)^n}} S(p-q)
\Gamma_\nu(p-q, q) G_\mu(k, q) G(k+q) D^0_{\mu\nu}(q),
\end{equation}
where $C_A$ is the quadratic Casimir operator in the adjoint representation.
Proceeding to the IR renormalized functions, we obtain

\begin{equation}
\bar B_1(p,k) =  {1\over 2} \bar g^2_1 Y  \bar S(p)
\bar \Gamma_\mu(p,0) \bar G_\mu(k,0) \bar G(k),
\end{equation}
iff a cancellation of the severe IR divergences takes place, i.e.,

\begin{equation}
\tilde{Z}_B^{-1} X(\epsilon) Z_2(\epsilon) Z_1^{-1}(\epsilon) \tilde{Z}_1 (\epsilon) \tilde{Z}_2
(\epsilon) = \epsilon Y, \qquad \epsilon \rightarrow  0^+,
\end{equation}
where $Y$ is an arbitrary but finite constant.
From the IR convergence condition (4.8) and the general ST identity IR convergence relation (4.5)
and Eq. (3.9), it follows

\begin{equation}
Y = Y_g.
\end{equation}
Substituting now the ghost SD equation (3.21) into the Eq. (4.7), on account of
the relation (4.9), one obtains

\begin{equation}
\bar B_1 (p, k) = - {1\over 2}i \bar S(p)
\bar \Gamma_\mu(p,0) \bar b (k^2) k_\mu.
\end{equation}
Let us note that this final expression does not depend explicitly on the
coupling constant as it should be.
It clearly shows that the first term of the $\bar B (p,k)$ skeleton expansion
is of order $k$ ($\sim O(k)$) as $k$ goes to zero
since $ \bar b(0)$ exists and is finite in this limit.

The analytical expression of the second skeleton diagram for the ghost-quark
scattering kernel $B(p,k)$ is

\begin{eqnarray}
&B_2(p,k)&= A g^4 \int {i d^n q \over {(2\pi)^n}} \int {i d^n l \over {(2\pi)^n}} S(p-q + l)
\Gamma_\beta (p-q+ l,l) S(p-q) \nonumber\\
&\Gamma_\nu(p-q, q)&G_\mu(k, -l) G(k-l) G_\alpha (k-l,q) G(k-l+q) D^0_{\alpha \nu}(q) D^0_{\mu \beta}(l),
\end{eqnarray}
where the constant $A$ is a result of the summation over color group indices
(its explicit expression is not important here, see below). As in the previous
case, by passing to the IR renormalized quantities and using twice the corresponding
IR convergent condition (4.8), we get

\begin{equation}
\bar B_2(p,k) = A_1 \bar g^4 Y^2 \tilde{Z}_B \bar S(p) \bar \Gamma_\mu (p,0)
\bar S(p) \bar \Gamma_\nu(p, 0)
\bar G_\mu(k,0) \bar G(k) \bar G_\nu (k,0) \bar G(k).
\end{equation}
Using further Eq. (3.21) again twice, we finally obtain

\begin{equation}
\bar B_2(p,k) = A_2 \tilde{Z}_B \bar S(p) \bar \Gamma_\mu (p,0)
\bar S(p) \bar \Gamma_\nu(p, 0) \bar b^2 (k^2) k_\mu k_\nu,
\end{equation}
which clearly shows that the second term is of order $k^2$ as $k$ goes to zero.

In the same way it is possible to show that the third term $\bar
B_3(p,k)$ of the skeleton expansion for the ghost-quark scattering
kernel $\bar B(p,k)$ is of order $k^3$ ($\sim O(k^3)$) as $k$ goes
to zero. These arguments are valid term by term in the skeleton
expansion for the ghost-quark scattering kernel. Thus we have the
estimate

\begin{equation}
\bar B_n(p,k) = O(k^n), \qquad k \rightarrow 0,
\end{equation}
which means that we can restrict ourselves to the first term in the skeleton
expansion of the $\bar B(p,k)$ kernel at small $k$, i.e, put

\begin{equation}
\bar B(p,k) = \bar B_1 (p, k) + O(k^2), \qquad k \rightarrow 0.
\end{equation}

Differentiating now the IR finite quark ST identity (4.4) with respect to $k_\mu$ and
passing to the limit $k=0$, we obtain

\begin{equation}
- i \bar \Gamma_\mu(p, 0) \left[ \tilde{Z}^{-1} (\epsilon)+
\bar b(0) \right] = \tilde{Z}_B^{-1} d_\mu \bar S^{-1} (p) -
 \bar \Psi_\mu (p) \bar S^{-1}(p) +  \bar S^{-1}(p) \bar \Psi_\mu (p),
\end{equation}
where

\begin{equation}
\bar \Psi_\mu (p) = \left[ { \partial \over \partial k_\mu } \bar B (p, k) \right]_{k=0}  = - {1\over 2}i \bar b (0)  \bar S(p) \bar \Gamma_\mu(p,0).
\end{equation}
Substituting the relation (4.17) back into the previous ST identity
(4.16), its IR renormalized version becomes

\begin{eqnarray}
\left[ \tilde{Z}^{-1}(\epsilon) + {1 \over 2} \bar b (0) \right] \bar \Gamma_\mu(p,0) = i \tilde{Z}_B^{-1} d_\mu \bar S^{-1}(p)
-{1\over 2} \bar b(0) \bar S(p) \bar \Gamma_\mu(p,0) \bar S^{-1}(p).
\end{eqnarray}

\subsection{Rescaling procedure}

At the first sight we have obtained a very undesirable result since the IR
renormalized ST identity (4.18) heavily depends on the arbitrary IRMR constants
which have no physical sense. It depends also on the arbitrary ghost self-energy at zero point. However, let us formulate
now a general method how to escape in the IR renormalized ST identity (4.18)
the explicit dependence on the arbitrary ghost self-energy at zero point
and the above-mentioned arbitrary IRMR constants. For this purpose,
let us rescale the vertex in the ST identity (4.18) in accordance with

\begin{equation}
 \tilde{Z}_B \left[ \tilde{Z}^{-1} (\epsilon) + {1 \over 2}
\bar b (0) \right] \bar \Gamma_\mu(p,0) \Longrightarrow \bar \Gamma_\mu(p,0).
\end{equation}
Then the ST identity (4.18) becomes

\begin{equation}
\bar \Gamma_\mu(p,0) = id_\mu \bar S^{-1}(p)
- (1 + \Delta)^{-1} \bar S(p) \bar \Gamma_\mu(p,0) \bar S^{-1}(p),
\end{equation}
where

\begin{equation}
\Delta = {2 \tilde{Z}^{-1} (\epsilon) \over \bar b(0)}.
\end{equation}
Let us note that the dependence on the auxiliary "IRMR constant"
$\tilde{Z}_B$
disappears as expected.
The only problem now is the behavior of the ghost self-energy IRMR constant
$\tilde{Z}(\epsilon)$ in the $\epsilon \rightarrow 0^+$ limit. As was
underlined in the preceding section, only two independent cases should be
considered.

1). The ghost self-energy IRMR constant
$\tilde{Z}(\epsilon)$ does not depend on $\epsilon$ at all, i.e., it is finite
but arbitrary, $\tilde{Z}(\epsilon) = \tilde{Z}= const$. In this case,
redefining the ghost self-energy at zero point in the IR renormalized ST
identity (4.20), one obtains

\begin{equation}
\bar \Gamma_\mu(p,0) = id_\mu \bar S^{-1}(p)
- b_1(0) \bar S(p) \bar \Gamma_\mu(p,0) \bar S^{-1}(p)
\end{equation}
and

\begin{equation}
b_1(0) = (1 + \Delta (0))^{-1} = ( 1 + [2 \tilde{Z}^{-1} / \bar b(0)])^{-1}.
\end{equation}
It is just the analogue of this identity in 4D QCD which was first
obtained
by Pagels in his pioneering paper on NP QCD [14]. Let us
formally consider $\Delta (0) = [2 \tilde{Z}^{-1} / \bar b(0)]$ as
small. Then expanding in powers of $\Delta$, one gets

\begin{equation}
(1 + \Delta (0))^{-1} = 1 - \delta =1 - \sum^{\infty}_{n=2} (-1)^n \Delta^{n-1}.
\end{equation}
Substituting this back into the previous ST identity, one finally obtains

\begin{equation}
\bar \Gamma_\mu(p,0) = id_\mu \bar S^{-1}(p) - \bar S(p) \bar \Gamma_\mu(p,0) \bar
S^{-1}(p) + \delta \bar S(p) \bar \Gamma_\mu(p,0) \bar S^{-1}(p),
\end{equation}
which makes it possible to take into account the arbitrary coefficient $b_1$ step by
step in powers of $\Delta$, starting from $\delta=0$. For the sake of simplicity,
in this approximation (to leading order, $\delta=0$)  this ST identity will
be used in what follows.

 2). The second available possibility is when the ghost self-energy IRMR constant
$\tilde{Z}(\epsilon)$ is singular as $\epsilon$ goes to zero, so its inverse vanishes
in this limit. In this case $\Delta =0$ identically (see Eq. (4.21)), and the
quark ST identity (4.20) finally becomes

\begin{equation}
 \bar \Gamma_\mu(p,0) = id_\mu \bar S^{-1}(p) - \bar S(p) \bar
\Gamma_\mu(p,0) \bar S^{-1}(p).
\end{equation}
It is just the analogue of this identity in 4D QCD which was obtained in our
investigation of NP QCD [18] (see also Ref. [10] and references therein).
It is automatically free from ghost complications
($\delta=0$ from the very beginning). At
the same time, it contains nontrivial information on quarks degrees of freedom
themselves provided by the quark-ghost sector (the second term in Eqs. (4.25)
and (4.26), while
the first term is, obviously, the standard WT-type contribution).

\section{Complete set of equations for the IR renormalized quark propagator }

The final system of equations obtained for the IR renormalized
quantities in the quark sector are presented by the quark SD equation (2.7)
and the quark ST identity (4.26), i.e.,

\begin{eqnarray}
S^{-1} (p) &=& Z_2(\epsilon) S_0^{-1} (p)+ \bar g^2  \Gamma_\mu(p,0) S(p) \gamma_\mu,
\nonumber \\
\Gamma_\mu(p,0) &=& id_\mu S^{-1}(p) - S(p) \Gamma_\mu(p,0) S^{-1}(p).
\end{eqnarray}
For simplicity here we removed "bars" from the definitions of the
IR renormalized Green's functions, retaining them only for the
coupling constant (which has the dimensions of mass) in order to
distinguish it from initial ("bare") coupling constant. It
contains all known finite numerical factors as well as the
rescaling factor from the previous section. The arbitrary but
finite constant $Y_q$ is put to unity without losing generality in
advance (see section VIII).

The Euclidean version of our parametrization
of the quark propagator is as follows:

\begin{equation}
i S(p) = \hat p A(p^2) - B(p^2),
\end{equation}
 so its inverse is

\begin{equation}
i S^{-1}(p) = \hat p \overline A(p^2) + \overline B(p^2)
\end{equation}
with

\begin{eqnarray}
\overline A(p^2) &=& A(p^2) E^{-1} (p^2), \nonumber\\
\overline B(p^2) &=& B(p^2) E^{-1} (p^2), \nonumber\\
E(p^2) &=& p^2 A^2(p^2) + B^2(p^2).
\end{eqnarray}
In order to solve the ST identity (the second of equations in the
system (5.1)), the simplest way is to represent the quark-gluon
vertex function at zero momentum transfer as its decomposition in
terms of four independent form factors, namely

\begin{equation}
\Gamma_{\mu} (p, 0) = \gamma_{\mu} F_1(p^2)+ p_{\mu} F_2(p^2) -
\hat p p_{\mu} F_3(p^2) - \hat p \gamma_{\mu} F_4(p^2).
\end{equation}
Substituting this representation into the second of Eqs. (5.1) and
doing some tedious algebra of the $\gamma$ matrices in 2D
Euclidean space, one obtains

\begin{eqnarray}
F_1(p^2) &=& - {1 \over 2} \overline A'(p^2), \nonumber\\
F_2(p^2) &=& - \overline B'(p^2) - F_4(p^2), \nonumber\\
F_3(p^2) &=& \overline A'(p^2), \nonumber\\
F_4 (p^2) &=& {1 \over 2} A(p^2) \overline A(p^2) B^{-1} (p^2),
\end{eqnarray}
where the prime denotes the derivative with respect to the
Euclidean momentum variable $p^2$.

It is convenient to introduce the dimensionless variables and
functions as

\begin{equation}
A(p^2) = \bar g^{-2} A(x), \qquad B(p^2) = \bar g^{-1} B(x), \qquad
 x = p^2/{\bar g^2}.
\end{equation}
Taking into  account the previous relations and definitions, and performing
further
the algebra of the $\gamma$ matrices in 2D Euclidean space, the
system (5.1) can be explicitly reduced to a system of a coupled, nonlinear
ordinary differential equations of the first order for the $A(x)$ and $B(x)$
quark propagator form factors.

\subsection{IR finite quark propagator}

For the quark propagator which is IR finite (IRF) from the very beginning,
i.e.,
when $Z_2(\epsilon) = Z_2 =1$ as $\epsilon$ goes to zero (see section VIII below),
 the system of equations (5.1) becomes

\begin{eqnarray}
S^{-1} (p) &=& S_0^{-1} (p)+ \bar g^2  \Gamma_\mu(p,0) S(p) \gamma_\mu,
\nonumber \\
\Gamma_\mu(p,0) &=& id_\mu S^{-1}(p) - S(p) \Gamma_\mu(p,0) S^{-1}(p).
\end{eqnarray}
Doing some of the above-mentioned tedious algebra, the quark SD
equation (5.8) is finally reduced to

\begin{eqnarray}
 x A' &=& - (1 + x) A - 1 - m_0 B, \nonumber\\
2B B' &=& - A^2  + 2 ( m_0 A - B)B,
\end{eqnarray}
 where $A \equiv A(x)$, $B \equiv B(x)$, and now
the prime denotes the derivative with respect to the Euclidean
dimensionless momentum variable $x$.
For the dimensionless current quark mass, we retain, obviously, the
same notation, i.e., $m_0 / \bar g \rightarrow m_0$.

 The exact solution of the system (5.9) for the dynamically generated
quark mass function is

\begin{equation}
 B^2(c, m_0; x) =  \exp(- 2x)
\int^c_x \exp(2x') \tilde{\nu}(x')\, dx' ,
\end{equation}
where $c$ is the constant of integration and

\begin{equation}
\tilde{\nu} (x) = A^2(x) +2 A(x) \nu(x)
\end{equation}
with

\begin{equation}
\nu (x) = - m_0 B(x) = xA'(x) + (1 + x)A(x) + 1.
\end{equation}
Then the equation determining the $A(x)$ function becomes

\begin{equation}
{d\nu^2 (x) \over dx}+ 2\nu^2(x)= - A^2(x) m^2_0 - 2 A(x) \nu(x) m_0^2.
\end{equation}

 In the chiral limit ($m_0 = 0$) the system (5.9) can be solved
exactly. The solution for the $A(x)$ function is

\begin{equation}
A_0(x) =  - x^{-1} \left\{ 1 - \exp(-x) \right\}.
\end{equation}
It has thus the correct asymptotic properties (is regular at small $x$ and
asymptotically approaches the free propagator at infinity). For the dynamically
generated quark mass function $B(x)$ the exact solution is

\begin{equation}
 B^2_0(c_0, x) =  \exp(- 2x)
\int^{c_0}_x \exp(2x') A^2(x')\, dx',
\end{equation}
where $c_0 = p_0^2 / \bar g^2$ is an arbitrary constant of
integration. It is regular at zero. In addition, it also has
algebraic branch points at $x=c_0$ and at infinity (at fixed
$c_0$). As in the general (nonchiral) case, these unphysical
singularities are caused by the inevitable ghost contributions in
the covariant gauges.

   As was mentioned above, $A_0(x)$ automatically has a correct
behavior at infinity (it does not depend on the constant of
integration since it was specified in order to get regular at zero
solution). In order to reproduce the correct behavior at infinity
($x \rightarrow \infty$) of the dynamically generated quark mass
function, it is necessary to pass simultaneously to the limit $c_0
\rightarrow \infty$ in Eq. (5.15). So it identically vanishes in
this limit in accordance with the vanishing current light quark
mass in the chiral limit. Obviously, we have to keep the constant
of integration $c_0$ in Eq. (5.15) arbitrary but finite in order
to obtain a regular at zero point solution. The problem is that if
$c_0 = \infty$, then the solution (5.15) does not exist at all at
any finite $x$, in particular at $x=0$.

Concluding, let us note that an exact solution which is singular
at zero also exists. It is easy to check that $A_0(x) = -(1/x)$
automatically satisfies the system (5.9) in the chiral limit. The
corresponding exact singular solution for the dynamically
generated quark mass function can be obtained by substituting this
expression into the Eq. (5.15).

\subsection{IR vanishing quark propagator}

For the IR vanishing (IRV) type of the quark propagator, when
$Z_2(\epsilon)$ vanishes as $\epsilon$ goes to zero, the final
system of equations (5.1) becomes

\begin{eqnarray}
\bar S^{-1} (p) &=& i \bar m_0 + \bar g^2  \bar \Gamma_\mu(p,0) \bar S(p) \gamma_\mu, \nonumber\\
\Gamma_\mu(p,0) &=& id_\mu S^{-1}(p) - S(p) \Gamma_\mu(p,0) S^{-1}(p),
\end{eqnarray}
where, obviously, $\bar m_0 = Z_2(\epsilon) m_0(\epsilon)$ exists
as $\epsilon$ goes to zero. Just this type of the quark propagator
in the light-cone gauge has been first investigated by 't Hooft
[1]. In terms of the dimensionless variables (5.7), similar to the
previous case, the system (5.16) can be reduced to

\begin{eqnarray}
 x A' &=& - A - 1 - \bar m_0 B, \nonumber\\
2B B' &=& - A^2  + 2 \bar m_0 A B,
\end{eqnarray}
 where again $A \equiv A(x)$, $B \equiv B(x)$, and
the prime denotes the derivative with respect to the Euclidean
dimensionless momentum variable $x$. For simplicity, we use the same
notation for the dimensionless current quark mass, i.e., $\bar m_0
/ \bar g \rightarrow \bar m_0$.

The exact solution of this system for the dynamically generated
quark mass function is

\begin{equation}
 B^2(c, \bar m_0; x) = \int^c_x  \tilde{\nu}(x')\, dx' ,
\end{equation}
where $c$ is the corresponding constant of integration and

\begin{equation}
\tilde{\nu} (x) = A^2(x) +2 A(x) \nu(x)
\end{equation}
with

\begin{equation}
\nu (x) = - \bar m_0 B(x) = xA'(x) + A(x) + 1.
\end{equation}
Then the equation determining the $A(x)$ function becomes

\begin{equation}
{d\nu^2 (x) \over dx} = - A^2(x) \bar m^2_0 - 2 A(x) \nu(x) \bar m_0^2.
\end{equation}

In the chiral limit ($\bar m_0=0$) exact solutions are

\begin{equation}
 A_0(x) = - 1 + {c'_0 \over x},
\end{equation}
and

\begin{equation}
 B^2_0(c_0, x) = \int^{c_0}_x  A^2(x')\, dx' ,
\end{equation}
where $c'_0$ and $c_0$ are the corresponding constants of
integration, respectively. Regularity at zero implies $c'_0 =0$,
so that one finally obtains

\begin{equation}
 A_0(x) = - 1, \quad B^2_0(c_0, x) = (c_0 - x),
\end{equation}
where we retain the same definition and notation as previously for
the constant of integration $c_0$. Again as in the previous case,
it should be kept finite (but it remains arbitrary) as well as a
simultaneous limit $x, c_0 \rightarrow \infty$ for the dynamically
generated quark mass function $B^2_0(c_0, x)$ (5.23) is required.

\section{Quark confinement}

In principle, it is possible to develop the calculation schemes in different
 modifications which give the solution of both systems (5.9) and (5.17) step by
step in powers of the light current quark masses as well as in the inverse
powers of the heavy quark messes.

The important observation, however, is that the formal exact
solutions (5.10) and (5.18) exhibit the algebraic branch point at
$x=c$ which completely $excludes \ the \ pole-type \ singularity$
at any finite point on the real axis in the $x$-complex plane
whatever the solution for the $A(x)$ function might be. Thus the
solution cannot be presented in either case as the expression
having finally a pole-type singularity at any finite point $p^2 =
- m^2$ (Euclidean signature), i.e.,

\begin{equation}
S(p) \neq {const \over \hat p + m},
\end{equation}
certainly satisfying thereby the first necessary condition of
quark confinement formulated at the fundamental quark level as the absence
of a pole-type singularity in the quark propagator [19].

In order to confirm this, let us assume the opposite to Eq. (6.1),
i.e., that is the quark propagator within our approach may have a
pole-type singularity like the electron propagator has in quantum
electrodynamics (QED) (see Eq. (6.4) below). In terms of the
dimensionless quark form factors, defined in Eq. (5.7), this means
that in the neighborhood of the assumed pole at $x = -m^2$
(Euclidean signature), they can be presented as follows:

\begin{eqnarray}
A(x) &=& { 1 \over (x + m^2)^{\alpha}} \tilde{A} (x), \nonumber\\
B(x) &=& { 1 \over (x + m^2)^{\beta}} \tilde{B} (x),
\end{eqnarray}
where $\tilde{A}(x)$ and $\tilde{B}(x)$ are regular at a pole,
while ${\alpha}$ and ${\beta}$ are in general arbitrary with $Re
{\alpha},{\beta} \geq 0$. However, substituting these expansions
into the systems (5.9) and (5.17) and analyzing them in the
neighborhood of the assumed pole, one can immediately conclude in
that the self-consistent systems for the quantities with tilde
exists iff

\begin{equation}
\alpha = \beta = 0,
\end{equation}
i.e., our systems (5.9) and (5.17) do not admit the pole-type
singularities in the quark propagator in complete agreement with
the above-mentioned.

This point deserves a more detail duscussion, indeed. The IR asymptotics
of the electron propagator in QED is [20] (Minkowski signature)

\begin{equation}
S(p) \sim {1 \over (p^2 - m^2)^{1 +\beta}},
\end{equation}
where $\beta = \alpha(\xi -3) / 2 \pi$ and here $\alpha$ is the
renormalized charge. Thus instead of a simple pole, it has a cut
whose strength can be varied by changing the gauge fixing
parameter $\xi$. However, there is, in general, the pole-type
singularity at the electrom mass $m$, indeed, i.e., in QED there
is no possibility, in general, to escape a pole-type singularity
in the electron Green's function. Contrast to QED, our general
solutions (5.10) and (5.18) have no pole-type singularities, only
the branch points at $x=c$. Not losing generality, one can put $c
= p^2_c / \bar g^2$ (different $p^2_c$ for different solutions, of
course), then it follows that at the branch point $p^2_c = p^2$
and this does not explicitly depend on $\xi$. At the same time, it
is obvious that the existence of a branch point itself does not
depend explicitly on a gauge choice as well. Thus the absence of
the pole-type singularities in QCD in
 the same way is gauge-invariant as the existence of the pole-type
 singularity at
the electron mass in QED. This may be used indeed to differentiate QCD
from QED and vice versa. The gauge invariance of the above-mentioned
first necessary condition of quark confinement should be precisely
understood in this sense.

Let us emphasize that the absence of the pole-type singularities
in the quark propagator as the criterion of confinement at the
fundamental quark level makes sense only for the $IR \
renormalized$ quark propagator, i.e., for entities having sense in
the $\epsilon \rightarrow 0^+$ limit. To speak about quark
confinement in the sense that the pole of the propagator is
shifted towards infinity as  $\epsilon \rightarrow 0^+$, and
therefore there is no physical single quark state, is though possible,
but confusing in our opinion (see Ref. [21] as well).
The problem is
that the quark propagator which is only IR regularized is not
physical, and so cannot be used to analyse such physical
phenomena as quark confinement, DCSB, etc.

The second sufficient condition formulated at the hadronic level as the
existence of a discrete spectrum only (no continuum in the spectrum) [1]
in
the bound-state problems within the corresponding BS formalism  is obviously
beyond the scope of the present investigation.
 Let us only note here, that at nonzero temperature the
bound-states will be dissolved (dehadronization), but the first necessary
condition of the quark confinement criterion will remain valid,
nevertheless. In other words, quarks at nonzero temperature (for example,
in the quark-gluon plasma (QGP) [22]) will remain off-shell objects, i.e.,
even in this case they cannot be detected as physical particles (like
electrons) in the asymptotic states. That is why it is better to speak
about dehadronization phase transition in QGP rather than about
deconfinement phase transition.

In both cases the region $c \ge x$ can be considered as NP
whereas the region $c \le x$ can be considered as the PT one.
Approximating the full gluon propagator by its free counterpart
 in the whole range $[ 0, \infty)$, nevertheless, we obtain a
solution for the dynamically generated quark mass function $B(x)$
which manifests the existence of the boundary value momentum
(dimensionless) $c$ (in the chiral limit $c_0$) separating the PT
region from the NP one, where the NP effects such as confinement
and DCSB become dominant. The arbitrary constant of integration
$c$ ($c_0$) is related to the characteristic mass which in 2D QCD
is nothing else but the coupling constant. So in 2D QCD (unlike 4D
QCD) there is no need to introduce explicitly into the quantum YM
theory the characteristic mass scale parameter, the so-called
Jaffe-Witten (JW) mass gap [23,24].

As was mentioned above, our solutions to the IR renormalized quark
propagator are valid in the whole momentum range $[ 0, \infty)$.
However, in order to calculate any physical observable from first
principles (represented by the corresponding correlation function
which can be expressed in terms of the quark propagator integrated
out), it is necessary to restrict ourselves to the integration
over the NP region $x \le c$ ($x \le c_0$) only. This guarantees
us that the above-mentioned unphysical singularity (branch-point
at $x=c$ ($x = c_0$)) will not affect the numerical values of the
physical quantities. Evidently, this is equivalent to the
subtraction of the contribution in the integration over the PT
region $x \ge c$ ($x \ge x_0$). Let us underline that at the
hadronic level this is the only subtraction which should be done
"by hand" (see discussion below in section XII, however) since our
solutions to the IR renormalized quark SD equations are
automatically NP. Thus there is no need for additional subtraction
of all types of the PT contributions at the fundamental
quark-gluon level in order to deal with the only true NP
quantities.
 In this connection, let us
remind the reader that many important quantities in QCD such as gluon and
quark condensates, topological susceptibility, etc. are defined beyond
the PT theory only [25,26]. This means that they are determined by such
$S$-matrix
elements (correlation functions) from which all types of the PT contributions
should be subtracted, by definition, indeed (see next section).

\section{Dynamical breakdown of chiral symmetry (DBCS)}

From a coupled systems of the differential equations (5.9) and
(5.17) it is easy to see that these systems (for the system (5.17)
the replacement $m_0 \rightarrow \bar m_0$ is assumed)

$allow \ a  \ chiral \ symmetry \ breaking \ solution \ only$,

\begin{equation}
m_0 = 0, \quad A(x) \ne 0, \ B(x) \ne 0
\end{equation}
and
$forbid \ a \ chiral \ symmetry \ preserving \ solution$,

\begin{equation}
m_0 = B(x) = 0, \quad A(x) \ne 0.
\end{equation}
Thus any nontrivial solutions automatically break the $\gamma_5$ invariance of
the quark propagator

\begin{equation}
\{ \gamma_5, S^{-1} (p) \} = - i\gamma_5 2 \overline B(p^2) \neq 0,
\end{equation}
and they therefore $certainly$ lead to the  spontaneous chiral symmetry breakdown at the fundamental quark level ($m_0 = 0, \ \overline{B}(x) \ne 0$,  dynamical quark mass generation).
In all previous investigations a chiral symmetry preserving solution (7.2)
always exists. For simplicity, we do not distinguish between $B(x)$ and
$\overline{B}(x)$
calling both dynamically generated quark mass functions.

A few preliminary remarks are in order. A nonzero, dynamically generated
quark
mass function defined by conditions (7.1) and (7.3) is the order parameter of
DBCS at the fundamental quark level. At the phenomenological level the order
parameter of DBCS is the nonzero quark condensate defined as the integral of
 the
trace of the quark propagator, i.e., (Euclidean signature, see Eq. (5.2))

\begin{equation}
< \bar q q> = <0|\bar q q|0> \sim i \int d^2p \ Tr S(p),
\end{equation}
up to unimportant (here and below in this section for our discussion) numerical
factors. In terms of the dimensionless variables (5.5) it becomes

\begin{equation}
< \bar q q>_0  \sim - \bar g \int dx \ B_0(x),
\end{equation}
where for light quarks in the chiral limit $< \bar q q>_0 =  < \bar u u>_0 =  < \bar d d>_0 =  < \bar s s>_0$, by definition, and integration over $x$ is
assumed from zero to infinity.

It is worth to emphasize now that the phenomenological order parameter of DBCS
- the quark condensate - defined as the dynamically generated quark mass
function $B(x)$ integrated out might be in principle zero even when the mass
function is definitely nonzero. Thus the nonzero, dynamically generated
quark
mass is a much more appropriate condition of DBCS than the quark condensate.
One can say that this is the first necessary condition of DBCS, while
the nonzero quark condensate is only the second sufficient one.

However, this  is not the whole story yet. The problem is that the quark
condensate defined in Eq. (7.5) still contains the contribution in the
 integration over the PT region, say,  $[y_0, \infty)$. In order to define
correctly the quark
condensate this contribution should be subtracted, i.e.,

\begin{equation}
< \bar q q>_0 \Longrightarrow < \bar q q>_0 + \bar g
\int_{y_0}^{\infty} dx \ B_0(x) = - \bar g \int_0^{y_0} dx \
B_0(x).
\end{equation}
If now the mass function $B(x)$ is really the NP solution of the
corresponding quark SD equation, then this definition gives the
quark condensate beyond the PT theory. In our case this is so,
indeed. Moreover, it is easy to understand that in order to
guarantee that the algebraic branch point at $x = c_0$ will not
affect the numerical value of the quark condensate,
 the soft cutoff $y_0$ should be identified with the constant
of integration $c_0$. Thus in our case it becomes

\begin{equation}
< \bar q q>_0 \sim - \bar g \int_0^{c_0} dx \ B_0(c_0, x),
\end{equation}
i.e., the truly NP dynamically generated quark mass function is
integrated out over the NP region as well. So there is not even a
bit of PT information in this definition (all types of the PT
contributions have been already subtracted in Eq. (7.7)).
Moreover, it depends on the fundamental mass scale parameter of 2D
QCD which is the IR renormalized coupling constant $\bar g$ and
not on the arbitrary mass scales $1 \ GeV$, $2 \ GeV$, etc. In the
PT limit $\bar g \rightarrow 0$ the quark condensate goes to zero
as it should be, by definition ($B_0(c_0, x)$ tends to zero in the
PT limit $c_0, x \rightarrow \infty$ as well). Thus in our
approach the quark condensate itself has a physical meaning, while
in other approaches, for example, in lattice QCD or in QCD sum
rules neither the quark condensate nor the current quark mass has
physical meaning by itself. Only they multiplication product gains
a physical sense becoming thus renormgroup invariant.
 In the same way the quark condensate should be defined in 4D
QCD though there is a problem with the JW mass gap as was mentioned above.

\section{IR multiplicative renormalizability of 2D QCD}

It is well known that 2D QCD is an UV, i.e., PT
super-renormalizable field theory [2,6]. However, the DT clearly
shows that this theory is IR divergent since its free gluon
propagator IR singularity is a NP (i.e., severe) one. For that
very reason, it becomes inevitable firstly to regularize it (which
has been already done), and then to prove its IR
renormalizability, i.e., to prove that all the NP IR singularities
can be removed from the theory on a general ground and in a
self-consistent way. In order to formulate the IRMR program in 2D
QCD, it is necessary to start from the quark-ghost sector.

\subsection{IRMR program in the quark-ghost sector}

 The IRMR program in the quark-ghost sector is based on the corresponding IR
convergence conditions: the quark SD condition (2.6), the ghost
self-energy condition (3.9), the quark ST identity condition (4.8)
and quark ST identity IR convergence relation (4.5). However, taking into
account the relation (4.9), only three of them are independent since by combining the ghost self-energy condition (3.9) with the general ST identity relation
(4.5), one obtains the IR convergence condition (4.8). Reminding the relation
(3.7) $\tilde{Z}_2(\epsilon) = \tilde{Z}^{-1}(\epsilon)$ and that
 the "IRMR constant" $\tilde{Z}_B$ does not depend on
 $\epsilon$, i.e., $\tilde{Z}_B(\epsilon)=\tilde{Z}_B$, the independent
 system of the IR convergence conditions can be
written as follows:

\begin{eqnarray}
X(\epsilon) Z^2_2(\epsilon) Z^{-1}_1(\epsilon) &=& \epsilon Y_q, \nonumber\\
X(\epsilon) \tilde{Z}_1(\epsilon) \tilde{Z}^{-2}(\epsilon) &=& \epsilon Y_g, \qquad  \epsilon \rightarrow  0^+, \nonumber\\
Z_1^{-1}(\epsilon) \tilde{Z} (\epsilon) &=& Z_2^{-1} (\epsilon) \tilde{Z}_B.
\end{eqnarray}
Thus, in general, we have five independent IRMR constants:
$X(\epsilon), \
Z_2 (\epsilon), \ Z_1(\epsilon), \ \tilde{Z}_1(\epsilon)$ and
$\tilde{Z}(\epsilon)$. We have also three arbitrary but finite constants
$Y_q, \ Y_g$ and $\tilde{Z}_B$. We know that the quark wave function
IRMR constant
$Z_2(\epsilon)$ cannot be singular, while the
ghost self-energy IRMR constant $\tilde{Z}(\epsilon)$ is either singular or
constant as $\epsilon \rightarrow  0^+$, otherwise all
the IRMR constants remain arbitrary in this limit.

Let us show now that the above-mentioned finite constants $Y_q, \
Y_g$ and $\tilde{Z}_B$ can be put to unity without losing
generality. For this purpose, let us redefine all the IRMR
constants as follows:

\begin{eqnarray}
X(\epsilon) &=& Y_q \tilde{Z}_B^{-2} X'(\epsilon), \quad  Z_2 (\epsilon)= \tilde{Z}_B Z'_2 (\epsilon), \quad Z_1(\epsilon) = Z'_1(\epsilon), \nonumber\\
\tilde{Z}_1(\epsilon) &=& Y_g Y_q^{-1} \tilde{Z}_B^2 \tilde{Z}'_1(\epsilon),
\quad  \tilde{Z}(\epsilon) = \tilde{Z}'(\epsilon).
\end{eqnarray}
Then it is easy to see that a new system for the IRMR
constants with primes looks like the previous system (8.1) if one puts there

\begin{equation}
\tilde{Z}_B = Y_q =Y_g =1.
\end{equation}
Thus in fact our system (8.1) is

\begin{eqnarray}
X(\epsilon) Z^2_2(\epsilon) Z^{-1}_1(\epsilon) &=& \epsilon , \nonumber\\
X(\epsilon) \tilde{Z}_1(\epsilon) \tilde{Z}^{-2}(\epsilon) &=& \epsilon, \qquad  \epsilon \rightarrow  0^+, \nonumber\\
Z_1^{-1}(\epsilon) \tilde{Z} (\epsilon) &=& Z_2^{-1} (\epsilon),
\end{eqnarray}
so we have three conditions for the above-mentioned five independent IRMR
constants. Obviously, this system has always a nontrivial solution
determining
three of the constants in terms of two arbitrary chosen independent IRMR
constants.
It is convenient to choose $\tilde{Z} (\epsilon)$ and $Z_2 (\epsilon)$ as two
 independent IRMR constants since we know their possible behavior
with respect to $\epsilon$ as it goes to zero. Then the general solution
of the
system (8.4) can be written as follows:

\begin{equation}
X(\epsilon) = \epsilon Z^{-1}_2(\epsilon) \tilde{Z}(\epsilon), \quad
Z_1(\epsilon) = \tilde{Z}_1(\epsilon) = Z_2 (\epsilon) \tilde{Z} (\epsilon).
\end{equation}

Thus in the quark-ghost sector the self-consistent IRMR program
really exists. Moreover, it has room for additional
specifications. The most interesting case is the quark
propagator which is IR
finite from the very beginning, i.e., when the
quark wave function IRMR constant $Z_2 (\epsilon) = Z_2 = const$.
In this case the system (8.4) becomes

\begin{eqnarray}
X(\epsilon) Z^{-1}_1(\epsilon) &=& \epsilon Z^{-2}_2 , \nonumber\\
X(\epsilon) \tilde{Z}_1(\epsilon) \tilde{Z}^{-2}(\epsilon) &=& \epsilon, \qquad  \epsilon \rightarrow  0^+, \nonumber\\
Z_1^{-1}(\epsilon) \tilde{Z} (\epsilon) &=& Z_2^{-1}.
\end{eqnarray}
Again, let us redefine all the IRMR constants
as follows:

\begin{equation}
X(\epsilon) = Z_2^{-1} X'(\epsilon), \ Z_1(\epsilon) =  Z_2 Z'_1(\epsilon), \
\tilde{Z}_1(\epsilon)
= Z_2 \tilde{Z}'_1(\epsilon), \ \tilde{Z}(\epsilon) = \tilde{Z}'(\epsilon).
\end{equation}
Then the system for quantities with primes will be the same as the
previous one,
putting there $Z_2=1$. This means that in fact our system (8.6) in this
case is

\begin{eqnarray}
X(\epsilon) Z^{-1}_1(\epsilon) &=& \epsilon, \nonumber\\
X(\epsilon) \tilde{Z}_1(\epsilon) \tilde{Z}^{-2}(\epsilon) &=& \epsilon, \qquad  \epsilon \rightarrow  0^+, \nonumber\\
Z_1^{-1}(\epsilon) \tilde{Z} (\epsilon) &=& 1,
\end{eqnarray}
which determines now four independent IRMR constants:
$X(\epsilon), \ Z_1(\epsilon), \ \tilde{Z}_1(\epsilon), \
\tilde{Z}(\epsilon)$. Its solution is

\begin{equation}
X(\epsilon) = \epsilon \tilde{Z} (\epsilon), \qquad  \tilde{Z}(\epsilon)
=  \tilde{Z}_1(\epsilon) = Z_1 (\epsilon),
\end{equation}
in complete agreement with the general solution (8.5).

It is worth to investigate in detail the case when
$\tilde{Z}(\epsilon) = K Z_2^{-1}(\epsilon)$, where $K$ is an arbitrary
 but finite
constant (see subsection C below). Then the general system (8.4) becomes

\begin{eqnarray}
X(\epsilon) Z^2_2(\epsilon) &=& \epsilon K, \nonumber\\
X(\epsilon) \tilde{Z}_1(\epsilon) Z^2_2(\epsilon) &=& \epsilon K^2, \qquad  \epsilon \rightarrow  0^+, \nonumber\\
Z_1^{-1}(\epsilon) &=& K^{-1}.
\end{eqnarray}
Let us, as before, redefine all the IRMR constants in
this system as follows:

\begin{equation}
X(\epsilon) = K^{-1} X'(\epsilon) \quad Z_1(\epsilon) =  K Z'_1(\epsilon),
\quad \tilde{Z}_1(\epsilon) = K \tilde{Z}'_1(\epsilon), \quad
Z_2(\epsilon) = K Z'_2(\epsilon).
\end{equation}
Then the system for quantities with primes will be the same as the
previous one,
putting there $K=1$. This means that in fact our system (8.10) is

\begin{eqnarray}
X(\epsilon) Z^2_2(\epsilon) &=& \epsilon, \nonumber\\
X(\epsilon) \tilde{Z}_1(\epsilon) Z^2_2(\epsilon) &=& \epsilon, \qquad  \epsilon \rightarrow  0^+, \nonumber\\
Z_1^{-1}(\epsilon) &=& 1.
\end{eqnarray}
Its solution is

\begin{equation}
X(\epsilon) = \epsilon Z_2^{-2}(\epsilon), \quad
\tilde{Z}(\epsilon) = Z_2^{-1} (\epsilon) = \tilde{Z}_2^{-1}(\epsilon),
 \quad \tilde{Z}_1(\epsilon) = Z_1^{-1} (\epsilon) = 1,
\end{equation}
again in complete agreement with the general solution (8.5). Evidently, the
system (8.12) is equivalent to the general system (8.4) if
one puts there the quark-gluon vertex IRMR constant to unity from the very
beginning, i.e., $Z_1^{-1}(\epsilon) = Z_1^{-1} = 1$.

The fact that all the arbitrary but finite constants can be put
equal to unity is a general feature of our IRMR program in the
quark-ghost sector which enables us to remove all the severe IR
divergences from the theory in a self-consistent way. This is
important, since otherwise these arbitrary but different, finite
constants having no physical meaning would ``contaminate'' the
equations of motion (see, for example, the quark SD equation
(2.7)). Concluding this subsection, let us emphasize once more
that in the quark-ghost sector the IRMR program is definitely
self-consistent.

\subsection{IR finite ST identities for pure gluon vertices}

In order to determine the IR finite bound-state problem within the BS
 formalism, it is necessary to know the IRMR constants of the three- and
 four-gluon
proper vertex functions which satisfy the corresponding ST identities
[6,27-31]. This information is also necessary to investigate the IR
properties
of all other SD equations in 2D QCD. It is convenient to start from the ST
identity for the three-gluon vertex [27,28]

\begin{eqnarray}
[1 + b(k^2)] k_{\lambda} T_{\lambda\mu\nu}(k, q, r) &=& d^{-1}(q^2) G_{\lambda\nu}(q,k)
 (g_{\lambda\mu} q^2 - q_{\lambda} q_{\mu}) \nonumber\\
&+& d^{-1}(r^2) G_{\lambda\mu}(r,k) (g_{\lambda\nu} r^2 - r_{\lambda} r_{\nu}),
\end{eqnarray}
where $k+q+r=0$ is assumed and $d^{-1}$ is the inverse of the exact
gluon form factor, while G's
are the corresponding ghost-gluon vertices (3.3). Let us now introduce
the IR renormalized triple gauge field proper vertex as follows:

\begin{equation}
T_{\lambda\mu\nu}(k, q, r) = Z_3(\epsilon) \bar T_{\lambda\mu\nu}(k, q, r),
\quad \epsilon \rightarrow 0^+,
\end{equation}
where $\bar T_{\lambda\mu\nu}(k, q, r)$ exists as $\epsilon$ goes to zero, by
definition. Passing to the IR renormalized quantities, one obtains

\begin{eqnarray}
[\tilde{Z}^{-1} (\epsilon) + \bar b(k^2)] k_{\lambda} \bar T_{\lambda\mu\nu}(k, q, r)
 &=& \bar G_{\lambda\nu}(q,k) d^{-1}(q^2)(g_{\lambda\mu} q^2 - q_{\lambda} q_{\mu}) \nonumber\\
&+& \bar G_{\lambda\mu}(r,k) d^{-1}(r^2)(g_{\lambda\nu} r^2 - r_{\lambda} r_{\nu}),
\end{eqnarray}
so that the following IR convergence relation holds

\begin{equation}
 Z_3(\epsilon) = \tilde{Z}^{-1} (\epsilon) \tilde{Z}_1 (\epsilon),
\quad \epsilon \rightarrow 0^+.
\end{equation}

Let us make a few remarks. Here and below we are considering
the inverse of
the free gluon propagator as IR finite from the very beginning,
 i.e., $d^{-1} \equiv \bar {d}^{-1}=1$. This is not a singularity at all
 and therefore it should not be treated as a distribution [9] (there is
 no integration over its momentum).

The corresponding ST identity for the quartic gauge field vertex
is [27,28]

\begin{eqnarray}
[1 + b(p^2)] p_{\lambda} T_{\lambda\mu\nu\delta}(p,q,r,s) &=& d^{-1}(q^2)
 (g_{\lambda\mu} q^2 - q_{\lambda} q_{\mu}) B^g_{\lambda\nu\delta}(q,p;r,s) \nonumber\\
&+& d^{-1}(r^2) (g_{\lambda\nu} r^2 -r_{\lambda}r_{\nu})
B^g_{\lambda\mu\delta}(r,p;q,s) \nonumber\\
&+& d^{-1}(s^2)(g_{\lambda\delta} s^2 - s_{\lambda} s_{\delta})B^g_{\lambda\mu\nu}(s,p;q,r)
 \nonumber\\
&-& T_{\mu\lambda\delta}(q,s,-q,-s) G_{\lambda\nu}(q+s,p,r) \nonumber\\
&-& T_{\mu\nu\lambda}(q,r,-q,-r) G_{\lambda\delta}(q+r,p,s) \nonumber\\
&-& T_{\nu\delta\lambda}(r,s,-r,-s) G_{\lambda\mu}(r+s,p,q),
\end{eqnarray}
where $p+q+r+s=0$ is assumed. Here T's and G's are the
corresponding three- and ghost-gluon vertices, respectively. The
quantity $B^g$ with three Dirac indices is the corresponding
ghost-gluon scattering kernel which is shown in Fig. 4 (see also Refs.
[6,29]).

\begin{figure}[bp]

\[ \input{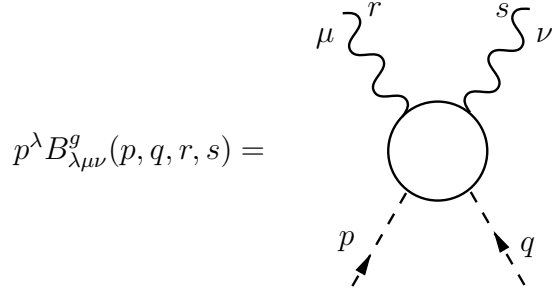}  \]

\caption{The ghost-gluon scattering kernel.}
\label{autonum}
\end{figure}

Let us introduce now its IR renormalized counterpart as follows:

\begin{equation}
B^g_{\lambda\nu\delta}(q,p; r,s) = \tilde{Z}_g(\epsilon) \bar
B^g_{\lambda\nu\delta}(q,p; r,s), \quad \epsilon \rightarrow 0^+,
\end{equation}
where $\bar B^g_{\lambda\nu\delta}(q,p; r,s)$ exists as $\epsilon$ goes to
zero. From the decomposition of the ghost-gluon proper vertex
shown in Fig. 5, it follows that

\begin{equation}
\tilde{Z}_1(\epsilon) = {1 \over \epsilon} X(\epsilon) \tilde{Z}_2 (\epsilon)
\tilde{Z}_g(\epsilon), \quad \epsilon \rightarrow 0^+,
\end{equation}
so that

\begin{equation}
\tilde{Z}_g(\epsilon) = \epsilon X^{-1}(\epsilon) \tilde{Z} (\epsilon)
\tilde{Z}_1(\epsilon), \quad \epsilon \rightarrow 0^+.
\end{equation}

\begin{figure}[bp]

\[ \input{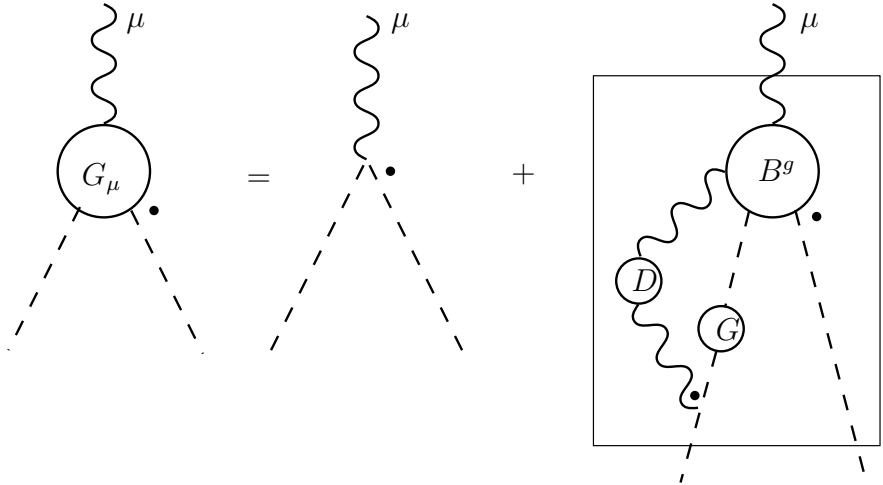}  \]

\caption{The decomposition of the ghost-gluon proper vertex.}
\label{autonum}
\end{figure}

It is worth reminding that to each ghost-gluon vertex a factor
$\sqrt{X(\epsilon)}$ should be additionally assigned, while to the
scattering kernel $B^g$ with two external gluon legs a factor
$X(\epsilon)$ should be additionally assigned.

Let us now introduce the IR renormalized four-gluon gauge field
vertex as follows:

\begin{equation}
T_{\lambda\mu\nu\delta}(p,q,r,s) = Z_4(\epsilon) \bar T_{\lambda\mu\nu\delta}(p,q,r,s),
\quad \epsilon \rightarrow 0^+,
\end{equation}
where $\bar T_{\lambda\mu\nu\delta}(p,q,r,s)$ exists as $\epsilon$ goes to zero, by definition.
 Passing again to the IR renormalized quantities, one obtains

\begin{eqnarray}
[\tilde{Z}^{-1}(\epsilon) + \bar b(p^2)] p_{\lambda} \bar
T_{\lambda\mu\nu\delta}(p,q,r,s) &=& d^{-1}(q^2)
 (g_{\lambda\mu} q^2 - q_{\lambda} q_{\mu}) \bar B^g_{\lambda\nu\delta}(q,p;r,s) \nonumber\\
&+& d^{-1}(r^2) (g_{\lambda\nu} r^2 -r_{\lambda}r_{\nu})
\bar B^g_{\lambda\mu\delta}(r,p;q,s) \nonumber\\
&+& d^{-1}(s^2)(g_{\lambda\delta} s^2 - s_{\lambda} s_{\delta}) \bar
B^g_{\lambda\mu\nu}(s,p;q,r)
 \nonumber\\
&-& \bar T_{\mu\lambda\delta}(q,s,-q,-s) \bar G_{\lambda\nu}(q+s,p,r) \nonumber\\
&-& \bar T_{\mu\nu\lambda}(q,r,-q,-r) \bar G_{\lambda\delta}(q+r,p,s) \nonumber\\
&-& \bar T_{\nu\delta\lambda}(r,s,-r,-s) \bar G_{\lambda\mu}(r+s,p,q),
\end{eqnarray}
iff

\begin{equation}
 Z_4(\epsilon) = Z_3^2(\epsilon) = \tilde{Z}^{-2} (\epsilon) \tilde{Z}_1^2 (\epsilon),
\quad \epsilon \rightarrow 0^+.
\end{equation}
Evidently, in the derivation of this expression the general solution (8.5) has
been  used as well as Eqs. (8.17) and (8.21). Thus we have determined the
IRMR constants of the triple and quartic gauge field vertices in Eqs. (8.17)
and (8.24), respectively.

\subsection{IR finite bound-state problem}

Apart from quark confinement and DBCS, the bound-state problem is
one of the most important NP problems in QCD. The general
formalism for considering it in quantum field theory is the BS
equation ([32,33] and references therein). For the color-singlet,
flavor-nonsinglet bound-state amplitudes for mesons it is shown in
Figs. 6 and 7.

\begin{figure}[bp]

\[ \input{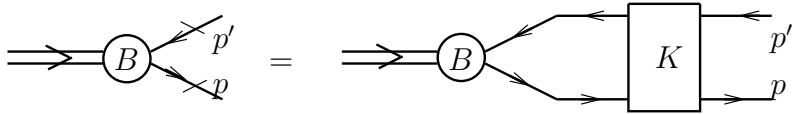}  \]

\caption{The BS equation for the flavored mesons.}
\label{autonum}
\end{figure}

\begin{figure}[bp]

\[ \input{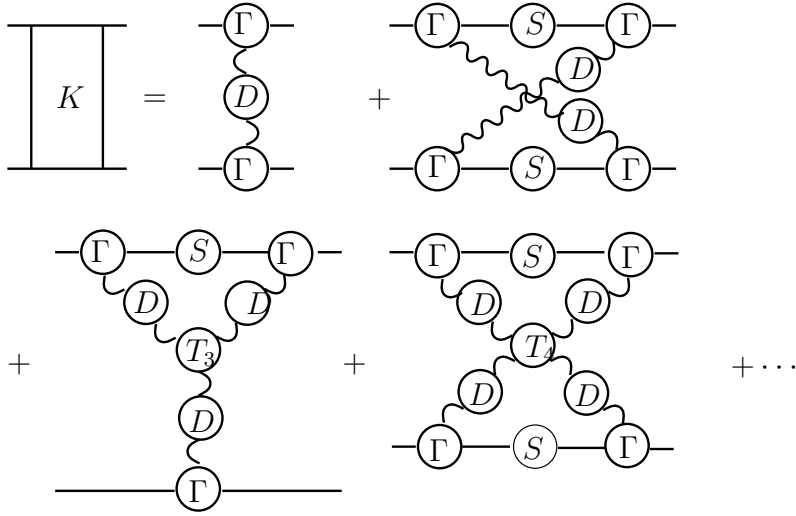}  \]

\caption{The skeleton expansion for the 2PI BS scattering kernel.}
\label{autonum}
\end{figure}

Flavor-singlet mesons require a special treatment since pairs,
etc. of gluons in color-singlet states can contribute to the
direct-channel processes. The exact BS equation for the
bound-state meson amplitude $B(p,p')$ can be written analytically
as follows (Euclidean signature):

\begin{equation}
S_q^{-1}(p)B(p,p')S_{\bar q}^{-1}(p') = i \int d^n l K(p,p'; l)
B(p,p'; l),
\end{equation}
(for simplicity all numerical factors are suppressed), where $S_q^{-1}(p)$ and $S_{\bar
q}^{-1}(p')$ are inverse quark and antiquark propagators, respectively, and $K(p,p';l)$
is the two-particle irreducible (2PI) BS scattering kernel (its skeleton
expansion is shown in
Fig. 7) which defines the BS equation itself. The BS equation is a homogeneous
linear integral equation for the $B(p,p')$ amplitude. For this reason the meson
bound-state amplitude should be always considered as IR finite from the
very beginning, i.e., $B(p,p') \equiv \bar B(p,p')$. Passing to the IR
renormalized  quantities in this equation, one obtains

\begin{equation}
\bar S_q^{-1}(p)B(p,p') \bar S_{\bar q}^{-1}(p') = i \int d^n l
\bar K(p,p'; l) B(p,p';l),
\end{equation}
iff

\begin{equation}
Z_2^{-2} (\epsilon) = Z_K(\epsilon), \quad \epsilon \rightarrow 0^+,
\end{equation}
where we introduce the IRMR constant $Z_K(\epsilon)$ of the BS scattering
kernel. This is the exact BS equation IR convergence condition.

In general, the nth skeleton diagram of the BS equation skeleton expansion
contains
$n$ independent loop integrations over the gluon momentum which
(as we already know) generates a
 factor $1 / \epsilon$ each, $n_1$ quark-gluon vertex
functions and $n_2$ quark propagators. Also it contains $n_3$ and $n_4$
three and
four-gluon vertices, respectively. It is worth reminding that to each
quark-gluon vertex and three-gluon vertex a factor $\sqrt{X(\epsilon)}$
should be additionally assigned, while to the four-gluon vertex a factor
$X(\epsilon)$ should be additionally assigned.
 Thus the corresponding IRMR constant is equal to

\begin{equation}
Z_K^{(n)}(\epsilon) = \epsilon^{-n} \Bigl[ Z_1^{-1}(\epsilon) \Bigr]^{n_1}
 \Bigl[ Z_2(\epsilon) \Bigr]^{n_2} \Bigl[ Z_3(\epsilon) \Bigr]^{n_3}
\Bigl[ Z_4(\epsilon) \Bigr]^{n_4} \Bigl[ X(\epsilon) \Bigr]^{n_4+ (n_3+n_1)/2}.
\end{equation}
On the other hand, it is easy to see that for each skeleton diagram the
following relations hold

\begin{equation}
n_2 = n_1 -2, \quad 2n = n_1 +n_3 + 2n_4.
\end{equation}
Substituting these relations into the previous expression and using the
general solution (8.5), as well as taking into account results of the
previous subsection, one finally obtains

\begin{equation}
Z_K^{(n)}(\epsilon) =  Z_2^{-2} (\epsilon) \Bigl[ Z_2(\epsilon) \tilde{Z}(\epsilon)
\Bigr]^{n -n_1},
\end{equation}
so that from Eq. (8.27) it follows that

\begin{equation}
\Bigl[ Z_2(\epsilon) \tilde{Z}(\epsilon) \Bigr]^{n-n_1} = A_{(n)},
\end{equation}
where $A_{(n)}$ is an arbitrary but finite constant different, in
principle, for each skeleton diagram. Evidently, its solution is

\begin{equation}
 \tilde{Z}(\epsilon) = \Bigl[ A_{(n)} \Bigr]^{{-1 \over n_1 - n}} Z_2^{-1}(\epsilon).
\end{equation}
Let us emphasize now that the relation between these (and all other) IRMR constants
cannot depend on the fact which skeleton diagram is considered. This means that
the
above-mentioned arbitrary but finite  constant must be a common factor for
all skeleton diagrams, i.e., $\Bigl[ A_{(n)} \Bigr]^{{-1 \over n_1 - n}} =
K$, where $K$ is again
arbitrary but finite and the solution becomes

\begin{equation}
 \tilde{Z}(\epsilon) = K Z_2^{-1}(\epsilon).
\end{equation}
However, we have already shown that all arbitrary but finite IRMR constants, in
particular this one (see relations (8.10)-(8.12)), should be put to unity not
losing generality.

Thus in order to determine the bound-state problem free from the IR
singularities within
the corresponding BS equation, the general solution (8.13) is relevant. This
 means
that we can forget about the ghost self-energy IRMR constant and have to
analyse everything in terms of the quark wave function IRMR constant
just because
of the relation (8.33) with $K=1$. When it goes to zero as $\epsilon \rightarrow 0^+$, then the
ghost self-energy IRMR constant is singular, while when $Z_2(\epsilon) = Z_2 =
const.$, then $\tilde{Z} (\epsilon)$ is also constant and, as we already know,
both constant can be put to one without losing generality.

\subsection{The general system of the IR convergence conditions}

The general system of the IR convergence conditions (8.13) for
removing at this stage all the severe IR singularities on a
general ground and in self-consistent way from the theory becomes

\begin{eqnarray}
X(\epsilon) &=& \epsilon Z^{-2}_2(\epsilon), \quad
\tilde{Z}^{-1}(\epsilon) = \tilde{Z}_2(\epsilon) = Z_3(\epsilon) =
\tilde{Z}_g(\epsilon) = Z_2(\epsilon), \nonumber\\
\tilde{Z}_1(\epsilon) &=& Z_1^{-1}(\epsilon) = 1, \quad Z_4(\epsilon) =
Z_3^2(\epsilon) = Z_2^2(\epsilon),
\end{eqnarray}
and the limit $\epsilon \rightarrow  0^+$ is always assumed. This
system provides the cancellation of all the severe IR
singularities in 2D QCD at this stage, and what is most important
this system provides the IR finite bound-state problem within our
approach. All the IRMR constants are expressed in terms of the
quark wave function IRMR constant $Z_2(\epsilon)$ except the
quark-gluon and ghost-gluon proper vertices IRMR constants. They
have been fixed to be unity though we were unable to investigate
the corresponding ST identity for the latter vertex (as was
mentioned in section III).

\section{IR finite SD equation for the gluon propagator}

Let us now investigate the IR properties of the SD equation for
the gluon propagator which is shown diagrammatically in Fig. 8 (see
also Refs. [34,35] and references therein). Analytically it can be
written as follows:

\begin{equation}
D^{-1}(q) = D^{-1}_0(q)  - {1 \over 2} T_t(q) -{1 \over
2} T_1(q) - {1 \over 2} T_2(q) - {1 \over 6} T'_2(q) + T_g(q) + T_q(q),
\end{equation}
where numerical factors are due to combinatorics and, for simplicity, the
Dirac indices determining
the tensor structure are omitted. $T_t$ (the so-called tadpole term)
and $T_1$ describe one-loop contributions, while $T_2$ and $T'_2$ describe two-loop
contributions containing three- and four-gluon proper vertices, respectively.
Evidently, $T_g, \ T_q$ describe ghost- and quark-loop contributions.

\begin{figure}[bp]

\[ \input{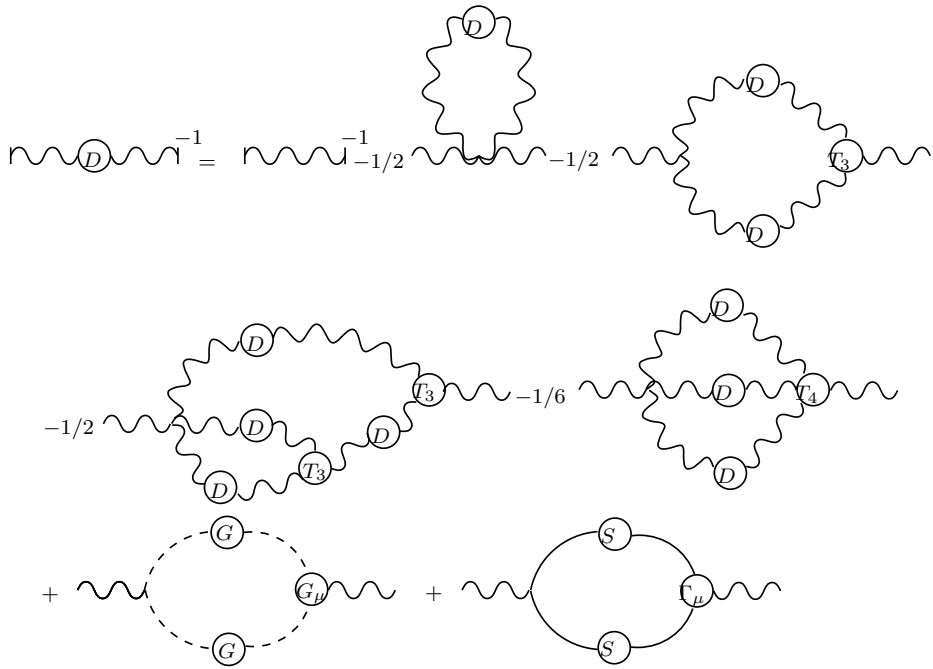}  \]

\caption{The SD equation for the gluon propagator.}
\label{autonum}
\end{figure}

Equating $D=D^0$ now and passing as usual to the IR renormalized
 quantities, one obtains

\begin{eqnarray}
{1 \over \epsilon} X(\epsilon) {1 \over 2} \bar T_t(q) &+& {1 \over \epsilon}
X(\epsilon) Z_3(\epsilon) {1 \over 2} \bar T_1(q) + {1 \over \epsilon^2}
X^2(\epsilon)Z_3^2(\epsilon) {1 \over 2} \bar T_2(q) + {1 \over \epsilon^2}
X^2(\epsilon)Z_4(\epsilon) {1 \over 6} \bar T'_2(q) \nonumber\\
&-& { 1 \over \epsilon} X(\epsilon)
\tilde{Z}^2_2 \tilde{Z}_1(\epsilon) \bar T_g(q) -  X(\epsilon) Z_2^2(\epsilon) Z_1^{-1}(\epsilon) \bar T_q(q) = 0,
\end{eqnarray}
where quantities with bar are, by definition, IR renormalized, i.e., they
exist as $\epsilon \rightarrow 0^+$. Let us also remind that each independent
loop integration over the gluon and ghost momenta generates the factor
$1 / \epsilon$,
while it is easy to show that there are no additional IR singularities with respect to $\epsilon$ in the quark loop (since we have found regular at zero solutions for the quark propagator). Using now the general solution (8.34), one
further obtains

\begin{equation}
{1 \over 2} \bar T_t(q) +Z_2(\epsilon) {1 \over 2} \bar T_1(q) + {1 \over 2} \bar
T_2(q) + {1 \over 6} \bar T'_2(q) - Z_2^2(\epsilon) \bar T_g(q)
- \epsilon Z_2^2(\epsilon) \bar T_q(q) = 0.
\end{equation}
Since the quark wave function IRMR constant $Z_2(\epsilon)$ can be only
either unity or
vanishing as $\epsilon$ goes to zero, the contribution from the
quark loop is always suppressed in the $\epsilon \rightarrow 0^+$ limit, and
we are left with
the pure YM SD equation for the gluon propagator. For the quark propagator
which is IR
renormalized from the very beginning (i.e., $Z_2(\epsilon) = Z_2 =1$, so that
it is IR finite), the SD equation (9.3) becomes

\begin{equation}
{1 \over 2} \bar T_t(q) + {1 \over 2} \bar T_1(q) + {1 \over 2} \bar T_2(q) +
{1 \over 6} \bar T'_2(q) = \bar T_g(q),
\end{equation}
while for the IR vanishing type of the quark propagator ($Z_2(\epsilon) \rightarrow 0$
as $\epsilon \rightarrow 0^+$), the SD equation (9.3) becomes

\begin{equation}
\bar T_t(q) + \bar T_2(q) + {1 \over 3} \bar T'_2(q) = 0.
\end{equation}

Concluding, let us note that it is not surprising that the IR properties
of the YM sector have been analyzed in terms of the quark wave function IRMR constant
$Z_2(\epsilon)$. Equivalently, it can be analyzed in terms of the ghost self-energy or
ghost propagator IRMR constants (because of the general solution (8.34)) which
are closely related to pure gluonic degrees of freedom via the corresponding ST
identities (see subsection B above). At the same time, the YM
SD equations for the gluon propagator (9.4) and (9.5) remain the
 same, of course. The tensor structure of the YM SD equations for the
gluon propagator is not important here. However, it may
substantially simplify the corresponding IR renormalized YM SD
Eqs. (9.4) and (9.5). Explicitly this should be done
elsewhere.\footnote{Let us remind that formally the $D=D^0$
solution always exists in the system of the SD equations due to
its construction by expansion around the free field vacuum [6]. It
is either trivial (coupling is zero) or nontrivial, then some
additional condition (constraint), including other Green's
functions, is to be derived. Eqs. (9.4) and (9.5) are precisely
these exact constraints.  The only question to be asked is whether
this solution is justified to use in order to explain some
physical phenomena, for example, quark confinement, DBCS, etc. or
not (for our conclusions see final section XII).}
 What matters here is that the self-consistent equations for the
gluon propagator free from the severe IR singularities exist in
the YM sector
 within our approach. In other words, the general
solution (8.34) eliminates all the severe IR singularities from
the Eq. (9.1), indeed.

\section{IR finite SD equation for the three-gluon proper vertex}

It is instructive to investigate the IR properties of the SD equation for the
triple gauge field proper vertex since it provides a golden opportunity to
fix $Z_2(\epsilon)$. This equation is shown in Fig. 9.

\begin{figure}[bp]

\[ \input{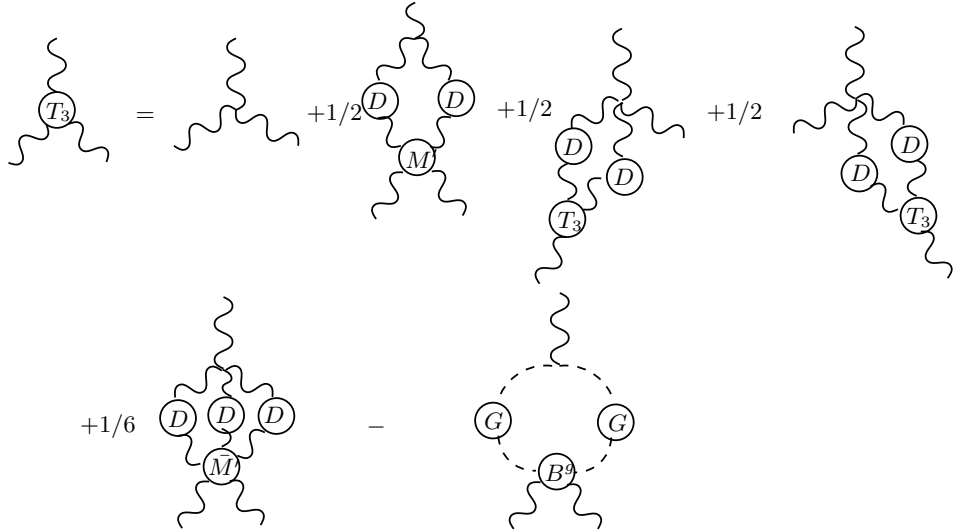}  \]

\caption{The SD equation for the triple gauge field vertex.}
\label{autonum}
\end{figure}

The skeleton expansions of the corresponding kernels are shown in Fig. 10.
Let us note that the ghost-gluon scattering kernel $B^g$ (for which we have
already established its IRMR constant from the decomposition of the ghost-gluon
proper vertex shown in Fig. 5 in subsection B, see also the general
solution (8.34)) is denoted as $G'$ in Ref. [6]. Obviously, there is no need
to investigate separately the IR properties of the SD equations for
the quark-gluon vertex and for pure gluon vertices since the information
about their IRMR constants has been uniquely extracted from the corresponding
ST identities. Moreover, the IRMR constants
of different types of the scattering kernels which enter the
above-mentioned SD equations (see, for example Figs. 8, 9 and 10) are to be
determined precisely by the general system (8.34). In principle, each
skeleton diagram of the above-mentioned expansions should be investigated
in the same way as was investigated the BS scattering kernel in subsection C.

 Since we know already the IRMR constant of the triple
gauge field proper vertex $Z_3(\epsilon)= Z_2(\epsilon)$, this makes it
possible to establish the IRMR constant of each scattering kernel
in general, i.e., not using its skeleton expansion.
For this purpose, let us apply
the same method which has been used in order to determine the IRMR constant
of the above-mentioned ghost-gluon scattering kernel. From the last term in
Fig. 9, it follows that

\begin{equation}
Z_3(\epsilon) = {1 \over \epsilon} X(\epsilon) \tilde{Z}^2_2(\epsilon)
\tilde{Z}_g(\epsilon)= \tilde{Z}_g(\epsilon)= Z_2(\epsilon),
\end{equation}
where in the second and third equalities the general solution (8.34) has been
used. Let us remind only that a factor
$X(\epsilon)$
should be additionally assigned to kernels with two gluon external legs,
while to the
kernels with three gluon external legs a factor $X^{3/2}(\epsilon)$
should be additionally assigned. Thus we
confirmed the result obtained earlier in subsection B for the IRMR constant
of the ghost-gluon scattering kernel $\tilde{Z}_g(\epsilon)$. Let us emphasize
that the left-hand side of the relation (10.1) should be equal to
$Z_3(\epsilon) = Z_2(\epsilon)$ since this skeleton diagram is nothing but
the corresponding independent decomposition of the triple gauge
field vertex itself.

However, the golden opportunity is provided by the third and fourth
terms of this SD equation. The interesting feature of these terms is that they
do not contain unknown scattering kernels, so their IR properties can be investigated directly by using only the known IRMR constants. On the other hand,
these terms are nothing but the corresponding
 decompositions of the triple gauge field proper vertex with the IRMR constant
is equal to $Z_3(\epsilon) = Z_2(\epsilon)$. Thus one has

\begin{equation}
Z_2(\epsilon) = {1 \over \epsilon} X(\epsilon)
Z_2(\epsilon) = Z_2^{-1}(\epsilon) = 1,
\end{equation}
which, obviously, has only a unique solution given by the last equality. Thus,
we have finally fixed the quark wave function IRMR constant to be unity.

\begin{figure}[bp]

\[ \input{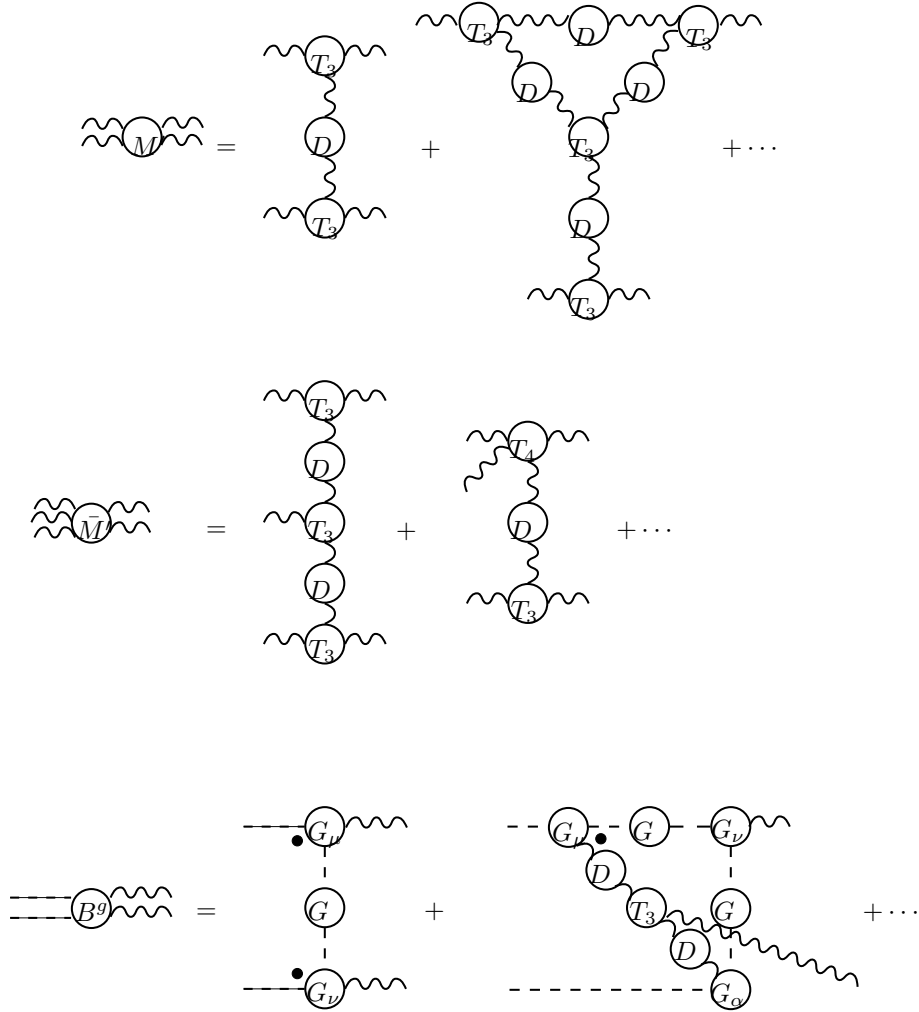}  \]

\caption{The skeleton expansions of different scattering kernels
in Fig. 9.}
\label{autonum}
\end{figure}

Adopting the same method, it is easy to show that all other IRMR
constants for the corresponding scattering kernels are

\begin{equation}
\tilde{Z}_g (\epsilon) = Z_{M'}(\epsilon) = Z_{\bar M'}(\epsilon) =1.
\end{equation}

We are now ready to investigate the IR properties of the SD
equation for the triple gauge field proper vertex shown in Fig. 9
without refereing to the skeleton expansions of the corresponding
scattering kernels (it is easy to check that the IRMR
constants of these kernels are consistent with their skeleton
expansions taking term by term). Using the previous results, the
IR renormalized version of this equation is

\begin{equation}
\bar T_3 = T_3^{(0)} + {1 \over 2} \bar T_1
+ {1 \over 2} \bar T'_1 + {1 \over 2} \bar T^{''}_1 + {1 \over 6} \bar T_2  -
\bar T_g,
\end{equation}
where, for simplicity, we omit the dependence on momenta and
suppress the Dirac indices (i.e., tensor structure) in all
terms in this SD equation. As usual the quantities
with bar are IR renormalized, i.e., they
exist as $\epsilon \rightarrow 0^+$.

\begin{figure}[bp]

\[ \input{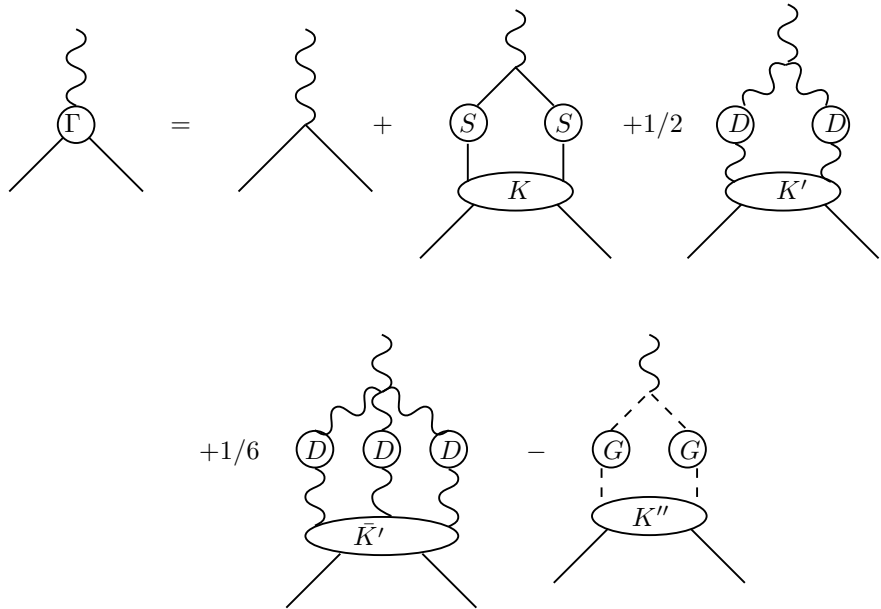}  \]

\caption{The SD equation for the quark-gluon proper vertex. The K's
are the corresponding scattering kernels.}
\label{autonum}
\end{figure}

Tthe SD equations for all other Green's functions can be investigated in
the same way, in particular for the quark-gluon
proper vertex shown in Fig. 11. The general solution (8.34), taking into
account the fundamental relation (10.2), provides their IR
convergence, i.e., they exist in the $\epsilon \rightarrow 0^+$
limit and, hence, similar to the SD equations, explicitly
considered here, they are free of
the IR divergences with respect to $\epsilon$.
Let us note that Eq. (9.5) should be ruled out as a possible SD equation
for the gluon propagator and the SD equation (9.4) is the
only possible one.

\section{Comparison with the 't Hooft model}

Having completed the proof of the IR renormalizability of our
approach to 2D QCD, it is instructive to compare it with the 't
Hooft model [1]. Of course, there is no direct comparison because
of the different gauges used. Nevertheless, one can still compare
the approximations made. It is well known that in the large $N_c$
limit and fixed $g^2N_c$ the quark loops are suppressed to the
leading order [1,8,36]. So the last term in the SD equation for
the gluon propagator (9.1), shown in Fig. 8, vanishes. Due to
light-cone gauge there are neither ghosts nor self-interaction of
massless gluons, and, therefore, the full gluon propagator becomes
equal to its free PT counterpart in this model (see Eq. (9.1)),
indeed. Thus from the whole of QCD only two sectors survive,
namely the quark and BS ones in the ladder approximation to the
quark-gluon proper vertex. Let us emphasize, nevertheless, that in
this model $D=D^0$ only to the leading order in the large $N_c$
limit, and nobody knows to what extent the next-to-leading order
corrections may distort the bahavior of the gluon propagator.

In order to reproduce the same approximation scheme within our approach, it is
necessary to neglect ghosts and the self-interaction of gluon fields "by hand",
while the quark loop contribution is automatically suppressed  as $\epsilon$ as
$\epsilon \rightarrow 0^+$ (see Eq. (9.3)). As a result, we are left
with only the quark and BS sectors which is absolutely similar to the 't
Hooft
model (though in the covariant gauge). For simplicity, here we are going to
discuss in some details the quark sector only.

It is easy to show that the SD equation for the quark propagator
(2.1) in the ladder approximation (the point-like quark-gluon
proper vertex) and within our treatment of the free gluon
propagator IR severe singularity, becomes

\begin{equation}
S^{-1}(p) = S_0^{-1}(p)+ {1 \over \epsilon} g^2 N_c \gamma_\mu S(p)\gamma_\mu,
\qquad \epsilon \rightarrow 0^+,
\end{equation}
where we include all known numerical factors into the coupling constant (having
the dimensions of mass) except of $N_c$. Let us remind that in the
't Hooft model [1] the IR regularization parameter was denoted as $\lambda$ and
in fact it was introduced by "hand" (though correctly). At the same time, in
this model it was assumed implicitly that the fixed combination $g^2N_c$
 is
IR finite from the very beginning (i.e., it does not depend on $\epsilon$ when
it goes to zero) though both parameters $g^2$ and $N_c$ (since it is the free
one in the large $N_c$ approach)
should, in general, depend on $\epsilon$ in the presence of such a strong
IR
singularity in the theory. There is also a possible problem of the commutation
of the two different limits: the IR limit, $\epsilon \rightarrow 0^+$ and the
large $N_c$ limit, $N_c \rightarrow \infty$. Anyway, in this model only the quark propagator becomes $\epsilon$-dependent, so the corresponding IR convergence
condition (2.8) in the quark sector is simply $Z^2_2(\epsilon) = \epsilon$ with
the obvious solution $\bar S(p) = \sqrt{\epsilon} S(p)$. In this way
one obtains for the IR renormalized quark propagator

\begin{equation}
\bar S^{-1} (p) = i\bar m_0+ g^2 N_c \gamma_\mu \bar S(p) \gamma_\mu,
\end{equation}
where $\bar m_0 = \sqrt{\epsilon}m_0(\epsilon)$ exists as
$\epsilon \rightarrow 0^+$, by definition, i.e., it is the IR renormalized
current quark mass. Using further our parametrization of the quark propagator
(5.2), (5.3) and doing a rather simple algebra of the $\gamma$
matrices in 2D Euclidean space, we finally obtain

\begin{equation}
B^{-1} + \bar m_0 = 2 g^2 N_c B, \qquad \bar A = A =0,
\end{equation}
i.e., the quark propagator has no $\gamma$ matrix structure, similar to
the 't Hooft model [1,2]. Moreover, the quark propagator is simply constant in
this approximation, namely

\begin{equation}
\bar S (p) = i B(p^2) = {i\bar m_0 \over 4g^2 N_c} \Bigl[ 1 \pm
\sqrt{1 + {8 g^2 N_c \over \bar m_0^2}} \Bigr].
\end{equation}
This means that in fact the current quark mass is replaced by the "effective
mass" $M$ as follows:

\begin{equation}
\bar m_0 \longrightarrow M = {4g^2 N_c \over \bar m_0 \Bigl[ 1 \pm
\sqrt{1 + {8 g^2 N_c \over \bar m_0^2}} \Bigr]},
\end{equation}
which in the chiral limit $\bar m_0 \rightarrow 0$ becomes

\begin{equation}
M_0^2 = 2 g^2 N_c.
\end{equation}

Using our formalism for the bound-state problem (subsection C), it
is easy to show that the same quark SD equation IR convergence
condition $Z^2_2(\epsilon) = \epsilon$ makes the BS sector IR
finite as well, i.e., free from the severe IR divergences. Thus
the 't Hooft model in the covariant gauge is almost trivially IR
renormalizable (as well as in the initial light-cone gauge).
Though the model quark propagator (11.4) is too simple,
nevertheless, its BS sector may be rather nontrivial, similar to
the BS sector of the initial 't Hooft model [1].

\section{Summary}

\subsection{Discussion}

In summary, the main observation is that 2D QCD is an inevitably
IR divergent theory. We have explicitly shown how the NP IRMR
program should be done in order to remove all the severe IR
singularities from the theory on a general ground and in a
self-consistent way. The general system of the IR convergence
conditions (8.34), taking into account the fundamental relation
(10.2), simply becomes

\begin{equation}
X(\epsilon) = \epsilon, \quad \epsilon \rightarrow  0^+,
\end{equation}
while all other independent quantities (Green's functions) are IR finite
from the very beginning, i.e., their IRMR constants are simply unity.
 Evidently,
only the nontrivial IR renormalization of the coupling constant is
needed to render the theory IR finite, i.e., to make it free from
all the severe IR divergences. Only the condition (12.1) provides
the cancellation of all the severe IR singularities in 2D
covariant gauge QCD. This completes the proof of the IR
renormalizability of 2D QCD within our approach.
 It is worth emphasizing once more that
it makes sense to discuss quark confinement, DBCS, the bound-state
problems, the tensor structure of the various SD rquations, etc.
only after the completion of the NP IRMR program, i.e., within
entities having sense as the IR regularization parameter goes
finally to zero (but not before).

Our proof implies that quark propagator should be IR finte from the very
beginning, i.e., $Z_2(\epsilon)=1$ which means $S(p) = \bar S(p)$.
In the 't Hooft model [1], the quark propagator is IR vanishing, i.e.,
$Z_2(\epsilon)$ goes to zero as $\epsilon \rightarrow 0^+$. However, there
is no contradiction with the above-mentioned since in this model neither
$g^2$ nor $N_c$ depend on $\epsilon$ (see section XI). From our general
solution (8.34) then it follows that $Z_2(\epsilon) = \sqrt{\epsilon}$,
indeed, since in this case one has to put $X(\epsilon)=1$.

One can conclude that in some sense it is easier to prove the IR
renormalizability of 2D QCD than to prove its UV
renormalizability. The reason is, of course, that we know the
mathematical theory which has to be used - the theory of
distributions [9]. This is due to its fundamental result [9,10]
which requires that any NP (severe) singularity with respect to
momentum in the deep IR domain in terms of $\epsilon$ should
always be $1 / \epsilon$ and this does not depend on how the IR
regularization parameter $\epsilon$ has been introduced in the way
compatible with the DT itself. On the other hand, the
above-mentioned fundamental result relates the IR regularization
to the number of space-time dimensions [9,10,23]
(compactification). It is easy to imagine that otherwise none of
the IRMR programs would be possible. In other words, the DT
provides the basis for the adequate mathematical investigation of
a global character of the severe (NP) IR divergences (each
skeleton independent loop diagram diverges as $1/ \epsilon$),
while the UV divergences have a local character, and thus should
be investigated term by term in powers of the coupling constant.

In this connection a few remarks are in order. The full dynamical
content of 2D (4D) QCD is contained in its system of the SD
equations of motion. To solve 2D (4D) QCD means to solve this
system and vice versa. In particular, to prove the IR
renormalizability of 2D (4D) QCD means to formulate the IRMR
program in order to remove all the NP IR singularities from this
system on a general ground and in a self-consistent way. As was
mentioned above, the fortunate feature which makes this possible
is a global character of the IR singularities in 2D QCD. Each
skeleton diagram is a sum of infinite series of terms, however,
the DT shows how their IR singularities can be summed up.
Moreover, it shows how the IR singularities of different
scattering kernels (which by themselves are infinite series of the
skeleton diagrams) can be summed up as well (see, for example,
sections VIII, IX, X).

 The next important step is to impose a
number of independent conditions in order to cancel all the NP IR
singularities which inevitably appear in the theory after the
above-mentioned summations have been done with the help of an
entire chain of strongly coupled SD equations. They should also be
complemented by the corresponding ST identities which are
consequences of the exact gauge invariance, and therefore are
exact constrains on any solution to QCD [6]. The only problem now
is to find a self-consistent solutions to the system of the IR
convergence conditions. If such solutions exist, so everything is
O.K. If not, the theory is not renormalizable. It is worth
reemphasizing that we have found a self-consistent solution to
this system (Eq. (12.1)).

Let us make a few remarks concerning the regularization and gauge
invariance of our approach. In principle, no regularization scheme
(how to introduce the IR regularization parameter in order to
parameterize the IR divergences) should be introduced "by hand".
First of all it should be well defined. Secondly, it should be
compatible with the DT [9]. The DR scheme [11] is precisely well
defined and  in Ref. [10] we have shown how it should be
introduced into the DT (complemented by the number of
subtractions, if necessary). The so-called "$\pm i\epsilon$
regularization" is equivalent to the regularization used in our
paper (see again Ref. [9]). Other regularizations schemes are also
available, for example, such as analytical regularization used in
Ref. [14] or the so-called Speer's regularization [37]. However,
they should be compatible with the DT as was emphasized above.
Anyway, not the regularization is important but the DT itself.

Whether the theory is IR multiplicative renormalizable or not
depends on neither the regularization nor the gauge. Due to the
chosen regularization scheme or the gauge only the details of the
corresponding IRMR program can be simplified. For example, in the
light-cone gauge at any chosen regularization scheme (the 't Hooft
model with different prescriptions how to deal with the severe IR
singularities [2] (and references therein)) to prove the IR
multiplicative renormalizability of 2D QCD is almost trivial. This
is mainly due to the fact that in this case only two sectors
survive in QCD, namely quark and BS sectors. In other words, if
theory is proven to be IR or UV renormalizable in one gauge, it is
IR or UV renormalizable in any other gauge. This is true for the
regularization schemes as well. As it follows from the present
investigation, to prove the IR multiplicative renormalizability of
2D QCD in the covariant gauge was not so simple. However, it was
necessary to get firstly the IR finite bound-state problem (which
is important for physical applications), and secondly to
generalize our approach on 4D QCD which is real theory of strong
interactions. 2D QCD in the light-cone gauge is not appropriate
theory for this purpose since its confinement mechanism looks more
like that of the Schwinger model [1] of 2D QED, than it may happen
in real QCD, where we believe it is much more complicated.

The structure of the severe IR singularities in Euclidean space is
much simpler than in Minkowski space, where kinematical
(unphysical) singularities due to light cone also exist. In this
case it is rather difficult to correctly untangle them from the
dynamical singularities, only ones which are important for the
calculation of any physical observable. Also the consideration is
much more complicated in configuration space [4]. That is why we
always prefer to work in momentum space (where propagators do not
depend explicitly on the number of dimensions)  with Euclidean
signature. We also prefer to work in the covariant gauges in order
to avoid peculiarities of the noncovariant gauges [38], for
example how to untangle the gauge pole from the dynamical one. The
IR structure of 2D QCD in the light-cone gauge by evaluating
different physical quantities has been investigated in more detail
in Refs. [2,21,39-41] (and references therein).

Of course, the quark propagator cannot be gauge-invariant
because the quark fields are not, by definition. This implicit gauge
dependence of the quark
propagator (as well as all other Green's functions) always exists and
cannot, in principle, be eliminated. This is a general feature of all
gauge theories such as QCD and QED.
Unfortunately, in gauge theories the main problem is not the
above-mentioned unavoidable implicit gauge dependence, but the explicit
dependence of the Green's functions on the gauge fixing parameter $\xi$
(on its numerical value). In the
quark SD equation it comes from the full gluon propagator and the
corresponding quark-gluon proper vertex. In both cases we have shown
that after the completion of our IRMR program (by correctly using the DT)
to remove all the NP IR divergences in a general way, the explicit gauge
dependence disappeared from the obtained system of eqautions (5.1).
This means that analytical properties of the solutions to this system
(the absence of the pole-type singularities and the presence of the
branch-points only) do not depend $explicitly$ on the gauge fixing
parameter, indeed.
Just in this sense the first necessary condition of the
quark confinement criterion discussed above is gauge-invariant.
Evidently, the second sufficient condition of quark confinement formulated
as the existence of a discrete spectrum (no continuum in the spectrum)
in the hadron spectroscopy is, by definition, gauge-invariant.

Also, it makes sense to bring the reader's attention to the
following point. The simplest approximation to the quark-gluon
vertex (compatible with the correct treatment of the IR
singularities by the DT in 2D QCD) is the vertex at zero momentum
transfer (see Eqs. (5.1)) and not its point-like counterpart. This
means that even in 2D QCD it is better to analyse confinement at
the fundamental quark level in terms of the analytical properties
of the quark
 propagator which reflect the IR structure of the 2D QCD true ground state.
 At the macroscopic, hadronic level the linear rising potential
interpretation of confinement becomes relevant for bound states
between heavy quarks only. In this case, apparently, the full
vertex can be approximated by its point-like counterpart, so the
analysis in terms of the potential becomes relevant. As was
mentioned above, that is why in the 't Hooft model [1] (where the
vertex is the point-like from the very beginning, see section XI)
confinement looks more like that of the Schwinger model [6] of 2D
QED. In real QCD it is believed to be much more complicated. This
complication is also due to nonabelian degrees of freedom, while
in the 't Hooft model they are eliminated by the choice of the
gauge.

\subsection{Conclusions}

We have shown that 2D covariant gauge QCD reveals several
desirable and promising features, so our main conclusions are:

1). We have proven the IR multiplicative renormalizability of 2D
QCD in the covariant gauge. It is based on the compelling
mathematical ground provided by the DT itself.

2). The nontrivial renormalization of the coupling constant only
makes theory free from all the severe IR singularities which
inevitably appear in 2D QCD.

3). The quark propagator has no poles, indeed (quark confinement).

4). It also implies DCSB, i.e., the chiral symmetry is certainly
dynamically (spontaneously) broken in 2D QCD.

5). We fixed finally the type of the quark propagator. The NP IRMR
program implies it to be IR finite from the very beginning as well
as all other Green's functions.

6). The bound-state problem becomes tractable within our our
approach. To any order in the skeleton expansion of the BS
scattering amplitude shown in Fig. 7, the corresponding BS
equation (8.26) can be reduced finally to an algebraic problem
(Appendix B).

7). The chiral limit physics (i.e., the Goldstone sector) can be
xactly evaluated since we have found exact solution for the quark
propagator in this case.

8). The nonzero quark masses can also easily included in our
scheme. We develop an analytical formalism which allows one to
find solution for the quark propagator in powers of the light
quark masses as well as in the inverse powers of the heavy quark
masses. We have theoretically justified the use of the free quark
propagator for heavy quarks. So our solution in this case
automatically possesses the heavy quark flavor symmetry (Appendix
A).

9). It was widely believed that the severe IR singularities couuld
not be put under control. However, we show explicitly that the
above-mentioned common belief is not justified. They be controlled
in all sectors of QCD of any dimensions by using correctly the DT
[9,10]. This can be considered also as one of our main results
from a mathematical point of view.

 10). We have proven that in order to accumulate the severe IR
singularities in 2D QCD, the YM SD equation for the gluon
propagator (9.4) is completely sufficient for this purpose.

The only dynamical mechanism responsible for quark confinement,
DBCS, the bound-states, etc. which can be thought of in 2D QCD is
the direct interaction of massless gluons. It becomes strongly
omain and can be effectively correctly absorbed into the gluon
propagator. It is well known that it is this interaction which
brings to birth asymptotic freedom (AF) [6] in QCD in the deep UV
limit. Thus the free gluon propagator due to its severe IR
structure is justified to use in order to explain all the
above-mentioned NP phenomena within our approach without
explicitly involving some extra degrees of freedom.

 A few points are worth reemphasizing as well.

The first important point is that the IR singularity of the free
gluon propagator, being strong at the same time, should be
correctly treated by the DT, complemented by the DR method. It
enables us to extract the required class of test functions in the
IR renormalized quark SD equation. The test functions do consist
of the quark propagator and the corresponding quark-gluon vertex
function. By performing the IRMR program, we have found the
regular solutions for the quark propagator. For that very reason
the relation (2.4) is justified since it is multiplied by the
appropriate smooth test functions [9]. Moreover, we establish the
space in which our generalized functions are continuous linear
functionals. It is a linear topological space denoted as $K(c)$
(for the solutions in the chiral limit denoted as $K(c_0)$),
consisting of infinitely differentiable functions having compact
support in $x \le c$ ($x \le c_0$), i.e., such functions which
vanish outside the interval $x \le c$ ($x \le c_0$) [9]. Thus the
above-mentioned subtraction of all kinds of the PT contributions
become not only physically well justified but well confirmed by
the DT (i.e., mathematically) as well.

The second point is that our theory (as mentioned above) is
defined by subtraction of all kinds of the PT contributions at the
fundamental quark level and at the hadronic level as well. The
only point of subtractions is the branch point. Thus we have exact
criterion how to separate the NP region (soft momenta) from the PT
region (hard momenta).

 The third point is that the system of
SD equations for the IR finite quantities (5.1) becomes
automatically free of the UV divergences though it is valid in the
whole momentum range $[0, \infty)$. At the same time, its
solutions, in general, and in the chiral limit, in particular,
preserve AF up to renormgroup log improvements, of course.

The fourth point is that the system of equations (5.1) for the IR
renormalized quark propagator is exact. Moreover, it does not
$explicitly$ depend on the gauge fixing parameter. It was obtained
in accordance with the rigorous rules of the DT, so there is no
place for theoretical uncertainties.

The fifth point is that the DT with its requirements to the
corresponding properties of the test functions removes all the
ambiguities from the theory. Because of this, all types of the
singular solutions should be excluded from the consideration, at
least in the standard DT sense.

\subsection{Some perspectives for 4D QCD}

We are not going here to evaluate the hadronic spectrum within our
approach. Anyway, it requires a separate treatment since, unlike
the 't Hooft model [1], our model is not simple (it cannot be
reduced to the ladder approximation in the BS sector). At the same
time, the bound-state problem becomes tractable within our
approach (see Appendix B).

 Our main concern is how to
generalize this approach on 4D QCD which is a realistic theory of
strong interactions not only at the fundamental quark-gluon level
but at the hadronic level as well. 2D covariant gauge QCD is a much more
appropriate theory to be generalized on 4D QCD than its 't Hooft
counterpart. Firstly, it maintains the direct interaction of massless
gluons (nonabelian degrees of freedom). Secondly, its dynamical structure
is much richer (full vertices, etc.). It is not accidental
that 2D light-cone gauge confinement mechanism at the fundamental
quark-gluon level turned out to be
almost useless to understand confinement mechanism in 4D QCD.

However, there are some principal distinctions between 2D and 4D
QCD. The most important one is that the former has initially the
JW mass gap which is the coupling constant itself. In the latter
case the coupling is dimensionless, so it is necessary to
introduce the JW mass gap from the very beginning into the quantum
4D YM theory. In close connection with this problem is the clear
understanding that the free gluon propagator is a bad
approximation to the full gluon propagator in the IR domain. Its
IR singularity is the PT one (i.e., not severe) in 4D QCD. So
necessarily the IR singularities of the full gluon propagator in
4D QCD should be stronger than $1/q^2$ as $q^2$ goes to zero. We
have already attempted to discuss both problems in more detail in
Ref. [24].

There still remains to resolve a set of some important problems.
Firstly, how to obtain the system of equations of motion in 4D QCD free from
possible strong IR singularities. We think that in this case the general IRMR
program should not be drastically different from that of 2D QCD formulated in
this work. Secondly, how to formulate the above-mentioned system of equations
of motion free from the explicit ghost degrees dependence and in a manifestly
 gauge invariant way, at least in the deep IR domain since there is no
hope for an exact solution(s). This is important for 4D QCD. Also
there should exist nontrivial PT dynamics in 4D QCD, while in 2D
QCD it is simple (we approximate the full gluon propagator by its
free PT counterpart in the whole momentum range). In this
approximation the 2D YM vacuum is also trivial (to all orders of
the vacuum-loops expansion [42]), while in 4D YM theory it can by
no means be trivial to any order. Anyway, the NP vacuum of 4D QCD
is expecting to be much more complicated. A generalization of our
approach to 2D and 4D QCD on nonzero temperature would be also
interesting.

\section{Acknowledgements}

The authors would like to thank L.P. Csernai and BPCL for their kind
support enabling us to carry out the substantial part of this work at the
Bergen University in the framework of the BPCL-17 Contract.
It is pleasure also to thank J. Revai and J. Nyiri for help. One of the
authors (V.G.) is grateful to A.V. Kouziouchine for help and support.

\appendix
\section{Nonzero quark masses}

To investigate solutions for the IR finite from the very beginning
quark propagator in the general (nonchiral case) it is much more
convenient to start from the ground system itself, Eqs. (5.9),
rather than to investigate the general solutions (5.10)-(5.13).
The ground system is

\begin{eqnarray}
 x A' + (1 + x) A + 1 &=& - m_0 B, \nonumber\\
2B B' + A^2 + 2 B^2 &=& 2 m_0 A B,
\end{eqnarray}
 where, let us remind, $A \equiv A(x)$, $B \equiv B(x)$, and
here the prime denotes the derivative with respect to the
Euclidean dimensionless momentum variable $x$ and the same
notation for the dimensionless current quark mass is retained
(i.e., $m_0 / \bar g \rightarrow m_0$). As was mentioned above, we
are interested in the solutions which are regular at zero and
asymptotically approach free quark case. Because of our
parametrization of the quark propagator (5.2) its asymptotic
behavior has to be determined as follows (Euclidean metrics):

\begin{eqnarray}
A (x) \sim_{x \rightarrow \infty} &-& { 1 \over x + m_0^2}, \nonumber\\
B (x) \sim_{x \rightarrow \infty} &-& { m_0 \over x + m_0^2},
\end{eqnarray}
up to renormgroup improvements by perturbative logarithms. The
ground system (A1) is very suitable for numerical calculations.

\subsection{Light quarks}

Let us now develop an analytical formalism which makes it possible
to find solution of the
 ground system step by step in powers of the light current quark masses, the so-called
 chiral perturbation theory at the fundamental quark level.
For this purpose let us present the quark propagator form factors
$A$ and $B$ as follows:

\begin{eqnarray}
A (x) &=& \sum_{n=0}^{\infty} m_0^n A_n (x), \nonumber\\
B (x) &=& \sum_{n=0}^{\infty} m_0^n B_n (x),
\end{eqnarray}
where

\begin{equation}
m_0^{(u,d,s)} \ll 1.
\end{equation}
Substituting these expansions into the ground system (A1) and
omitting some tedious algebra, one finally obtains

\begin{eqnarray}
 x A'_0(x) + (1 + x) A_0(x) + 1 &=& 0 , \nonumber\\
2B_0(x) B'_0(x)+ A^2_0(x) + 2 B^2_0(x) &=& 0,
\end{eqnarray}
and for $n=1,2,3,...$, one has

\begin{eqnarray}
x A'_n(x) + (1 + x) A_n(x) &=& - B_{n-1}(x) , \nonumber\\
2P_n(x) + M_n(x) + 2 Q_n(x) &=& 2 N_{n-1}(x),
\end{eqnarray}
where

\begin{eqnarray}
P_n (x) &=& \sum_{m=0}^n B_{n-m} (x) B'_m (x), \nonumber\\
M_n (x) &=& \sum_{m=0}^n A_{n-m} (x) A_m (x), \nonumber\\
Q_n (x) &=& \sum_{m=0}^n B_{n-m} (x) B_m (x), \nonumber\\
N_n (x) &=& \sum_{m=0}^n A_{n-m} (x) B_m (x).
\end{eqnarray}
Is is obvious that the system (A5) describes the ground system
(A1) in the chiral limit ($m_0=0$). As we already know it can be
solved exactly (see below as well). The first nontrivial
correction in powers of a small $m_0$ is determined by the
following system which follows from Eqs. (A6) and it is

\begin{eqnarray}
x A'_1 + (1 + x) A_1 &=& - B_0 , \nonumber\\
(B_1B'_0 + B_0 B'_1) +  A_0 A_1  + 2 B_0 B_1 &=& A_0 B_0,
\end{eqnarray}
where we omit the dependence on the argument $x$ for simplicity.
In the similar way can be found the system of equations to
determine terms of order $m_0^2$ in the solution for the quark
propagator and so on.

Let us present a general solution to the first of Eqs. (A6) which
is

\begin{equation}
 A_n(x) = - x^{-1} e^{-x} \int_0^x dx' \ e^{x'} B_{n-1} (x').
\end{equation}
It is always regular at zero since all $B_n(x)$ are regular as
well.
 The advantage of the developed chiral perturbation theory at the fundamental
 quark level is that each correction in the powers of small currect quark
 masses is determind by the corresponding system of equations which can be
 formally solved exactly.

The differential equation (A8) which determines first correction
for the dynamucally generated quark mass function is

\begin{equation}
B'_1 + [1 - {1 \over 2} A_0^2B_0^{-2}] B_1 =A_0 - A_0A_1B^{-1}_0 ,
\end{equation}
where we used the second of Eqs. (A5). It is easy to check that
its solution which is regular at zero is

\begin{equation}
 B_1(x) = \mu^{-1}(x) \int_{c_1}^x dz \ A_0(z)[1 -A_1(z)B_0^{-1}(z)]
\mu(z),
\end{equation}
where

\begin{equation}
\mu (x) = \exp [ x - {1 \over 2} a(x)],
\end{equation}
and

\begin{equation}
a(x) = \int_0^x dx' \ A_0^2(x')B_0^{-2}(x').
\end{equation}

Let us write down the system of solutions approximating the light
quark propagator up to first corrections, i.e.,

\begin{eqnarray}
A(x) &=& A_0(x) + m_0 A_1(x) + ...., \nonumber\\
B(x) &=& B_0(x) + m_0 B_1(x) + ....
\end{eqnarray}
This system is

\begin{equation}
A_0 (x) = - x^{-1} (1 - e^{-x}), \quad A_0(0) = -1 ,
\end{equation}

\begin{equation}
 B^2_0(x) = e^{-2x} \int_x^{c_0} dx' \  e^{2x'} A_0^2 (x').
\end{equation}
 And

\begin{equation}
 A_1(x) = - x^{-1} e^{-x} \int_0^x dx' \ e^{x'} B_0 (x'),
\end{equation}

\begin{equation}
 B_1(x) = e^{-2x}B_0^{-1}(x) \int_{c_1}^x dz \ e^{2z} A_0(z)[B_0(z)
 -A_1(z)],
\end{equation}
where again we use second of Eq. (A5) in order to integrated out
the $\mu(x)$ function. In physical applications we also need
$B^2(x)$, so we have

\begin{eqnarray}
B^2(x) &=& B_0^2(x) + 2 m_0 B_0(x) B_1(x) + ... \nonumber\\
&=& B_0^2(x) + 2 m_0 e^{-2x} \int_{c_1}^x dz \ e^{2z}
A_0(z)[B_0(z) -A_1(z)]  + ...,
\end{eqnarray}
and the relation between constants of integration $c_0$ and $c_1$
remains, in general, arbitrary. However, there exists a general
restriction, namely $B^2(x) \geq 0$ and it should be real which
may lead to some bounds for the constants of integration, while $x
\leq c_0$ always remains.

\subsection{Heavy quarks}

For heavy quarks it makes sense to replace $m_0 \rightarrow m_Q$.
In this case it is convenient to find solution for heavy quark
form factors $A$ and $B$ as follows:

\begin{eqnarray}
m_Q^2A (x) &=& \sum_{n=0}^{\infty} m_Q^{-n} A_n (x), \nonumber\\
m_Q B (x) &=& \sum_{n=0}^{\infty} m_Q^{-n} B_n (x),
\end{eqnarray}
and for heavy quark masses we have

\begin{equation}
m_Q^{(c,b,t)} \gg 1,
\end{equation}
i.e., the inverse powers are small. Substituting these expansions
into the first equation of the ground system (A1) and omitting
some tedious algebra, one finally obtains

\begin{eqnarray}
B_0(x) &=& -1, \nonumber\\
B_1(x) &=& 0, \nonumber\\
\end{eqnarray}
and

\begin{equation}
xA'_n(x) + (1 + x)A_n (x) = - B_{n+2}(x), \quad n= 0,1,2,3,...
\end{equation}
In the same way, by equating terms at equal powers in the inverse
of heavy quark messes, from second of the equations of the ground
system, one finally obtains

\begin{eqnarray}
P_0(x) + Q_0(x) - N_0(x) &=& 0, \nonumber\\
P_1(x) + Q_1(x) - N_1(x) &=& 0.
\end{eqnarray}
and

\begin{equation}
P_{n+2}(x) + Q_{n+2}(x) - N_{n+2}(x) = -{1 \over 2}M_n(x),
  \quad n=0,1,2,3,...,
\end{equation}
where $P_n(z), \ M_n(z), \ Q_n(z), \ N_n(z)$ are again given by
Eqs. (A7). Solving these equations, one obtains

\begin{eqnarray}
A_0(x) &=& B_0(x)= -1, \nonumber\\
A_1(x) &=& B_1(x) = 0,
\end{eqnarray}
and

\begin{eqnarray}
x A'_n(x) + (1+x) A_n(x) &=& - B_{n+2}(x), \nonumber\\
P_{n+2}(x) + Q_{n+2}(x) - N_{n+2}(x) &=& - {1 \over 2} M_n(x),
\quad n = 0,1,2,3, ...
\end{eqnarray}
It is possible to show that all odd terms are simply zero, i.e.,

\begin{equation}
A_{2n+1}(x)= B_{2n+1}(x) = 0,
 \quad n = 0,1,2,3, ...
\end{equation}
The explicit solutions for a few first nonzero terms are

\begin{equation}
A_0(x)= B_0(x) = - 1.
\end{equation}

\begin{eqnarray}
A_2(x) &=& x + {3 \over 2}, \nonumber\\
B_2(x) &=& x + 1.
\end{eqnarray}

\begin{eqnarray}
A_4(x) &=& - x^2 - {3 \over 2} x - {15 \over 2}, \nonumber\\
B_4(x) &=& - x^2 - {7 \over 2} x - {3 \over 2}.
\end{eqnarray}
Thus our solutions for the heavy quark form factors look like

\begin{eqnarray}
A (x) &=& { 1 \over m_Q^2} \sum_{n=0}^{\infty} m_Q^{-n} A_n (x) \nonumber\\
 &=& - {1 \over m_Q^2} + { x \over m_Q^4} - {x^2 \over m_Q^6} + ...+  D_A (x),
\end{eqnarray}
where

\begin{equation}
D_A(x) = { 3 \over 2 m_Q^4} - { 3x + 15 \over 2 m_Q^6} + ...
\end{equation}
And

\begin{eqnarray}
B(x) &=& { 1 \over m_Q} \sum_{n=0}^{\infty} m_Q^{-n} B_n (x) \nonumber\\
 &=& - {1 \over m_Q} + { x \over m_Q^3} - {x^2 \over m_Q^5} + ...+  D_B (x),
\end{eqnarray}
where

\begin{equation}
D_B(x) = { 1 \over  m_Q^3} - {7 x + 3 \over 2 m_Q^5} + ...
\end{equation}
Summing up, one obtains

\begin{eqnarray}
A (x) &=& - { 1 \over x + m_Q^2} + D_A (x), \nonumber\\
B(x) &=& - {m_Q \over x+ m_Q^2} +  D_B (x).
\end{eqnarray}

In terms of the Euclidean dimensionless variables (5.7), the heavy
quark propagator (5.2) is

\begin{equation}
iS(x) = \hat x A(x) - B(x).
\end{equation}
 Using our solutions, obtained above, it can be written down as
follows:

\begin{equation}
iS(x) = \hat x \Bigl( - { 1 \over x + m_Q^2} + D_A(x) \Bigr) +
{m_Q \over x + m_Q^2} - D_B(x).
\end{equation}
In other words, it becomes

\begin{equation}
iS(x) = iS_0(x) + \hat x D_A(x)  - D_B(x),
\end{equation}
where $iS_0(x)$ is nothing else but the free quark propagator with
the substitution $m_0 \rightarrow m_Q$, i.e.,

\begin{equation}
iS_0(x) = - { \hat x - m_Q \over x + m_Q^2}.
\end{equation}
Since $\hat x D_A(x)$ and $D_B(x)$ both are of the same order in
the inverse powers of $m_Q$, namely they are of order $m_Q^{-3}$,
 then Eq. (A39), becomes

\begin{equation}
iS(x) = iS_0(x) + 0(m_Q^{-3}).
\end{equation}
This means that our solution for the heavy quark propagator is
reduced to the free quark propagator up to terms of order $1 /
m_Q^3$.

\subsection{Heavy quarks flavor symmetry}

Let us explicitly show here that our solutions (A36) possesse the
heavy quark flavor symmetry [43]. We will show that the quark
propagator to leading order in the inverse powers of the heavy
quark mass will not depend on it, i.e., it is manifestly flavor
independent to the leading order of this expansion. For this
purpose, we must take into account that argument $x$ which is the
dimensionless momentum of the heavy quark contains itself the
heavy quark mass $m_Q$. In other words, a standard heavy quark
momentum decomposition should be used, namely

\begin{equation}
p_{\mu} = m_Q \upsilon_{\mu} + k_{\mu},
\end{equation}
as well as

\begin{equation}
\hat x = \gamma_{\mu} x_{\mu} = \gamma_{\mu} (m_Q \upsilon_{\mu} +
y_{\mu}),
\end{equation}
where $\upsilon$ is the four-velocity with $\upsilon^2=-1$
(Euclidean signature). It should be identified with the
four-velocity of the hadron. The "residual" momentum $k$ is of
dynamical origin. In these terms the Euclidean dimensionless
dynamical momentum variable $x=p^2/ \bar g^2$ then becomes

\begin{equation}
x = - m_Q^2 - 2 m_Q t - z,
\end{equation}
where we denote $t = (\upsilon \cdot y)$ with $y_{\mu} = k_{\mu} /
\bar g$ and $z= k^2/ \bar g^2$.

 Substituting expressions (A43) and (A44) and taking into account
only leading order terms in the inverse powers of $m_Q$, one
finally obtains

\begin{equation}
iS_h(\upsilon, y) = iS_0(\upsilon, y) + O({1 \over m_Q}),
\end{equation}
where

\begin{equation}
iS_0(\upsilon, y) = {1 \over \upsilon \cdot y } {\hat \upsilon  -
1 \over 2},
\end{equation}
which is exactly the heavy quark propagator [43]. Thus our
propagator does not depend on $m_Q$ to leading order in the heavy
quark mass limit, $m_Q \rightarrow \infty$, i.e., in this limit it
possesses the heavy quark flavor symmetry, indeed.

Concluding, let us note that the general system (A1) does not
demonstrate the principal difference in the analytical structure
of its solutions for light and heavy quarks. Also at the
fundamental quark level the heavy quark mass limit is not Lorentz
covariant. That is why in the case of heavy quarks we will use
rather Eq. (A39) than Eq. (A45).

\section{Bound-state problem}

Here let us only schematically show that the BS equation within
our approach can be reduced to an algebraic problem, indeed. The
BS equation (8.25) for the bound-state meson amplitude $B(p,p')$
to leading order (first skeleton diagram in Fig. 7) in the
skeleton expansion of the 2PI BS scattering kernel can be written
analytically as follows (Euclidean signature):

\begin{equation}
S_q^{-1}(p)B(p,p')S_{\bar q}^{-1}(p') =- i g^2 \int d^2l
\Gamma_{\mu}(p',l) B(p,p';l) \Gamma_{\nu}(p,l) D^0_{\mu\nu}(l),
\end{equation}
(for simplicity all numerical factors are suppressed), where as
usual $S_q^{-1}(p)$ and $S_{\bar q}^{-1}(p')$ are inverse quark
and antiquark propagators, respectively. Proceeding absolutely in
the same way as in section II, on account of the Laurent expansion
(2.4), one finally gets

\begin{equation}
S_q^{-1}(p)B(p,p')S_{\bar q}^{-1}(p') = { 1 \over \epsilon} g^2
\Gamma_{\mu}(p',0) B(p,p') \Gamma_{\mu}(p,0),
\end{equation}
where all numerical factors again were included into the coupling
constant. It is already known that the renormalization of the
coupling constant only is needed to get theory IR finite, i.e.,
$g^2 = X(\epsilon) \bar g^2$. Taking now into account the relation
(12.1), the IR renormalized bound-state problem becomes

\begin{equation}
S_q^{-1}(p)B(p,p')S_{\bar q}^{-1}(p') = \bar g^2
\Gamma_{\mu}(p',0) B(p,p') \Gamma_{\mu}(p,0).
\end{equation}
Thus we came to an algebraic problem to solve, indeed. It is
necessary to remind that the explicit solution for the quark-gluon
vertex at zero momentum transfer is given in Eqs. (5.5) and (5.6).
The bound-state amplitude $B(p,p')$, for example, for the
pseudoscalar meson-quark-antiquark vertex function is

\begin{equation}
G^i_5(p'+q,p') = \Bigl( {\lambda^i \over 2} \Bigr)  \gamma_5 [ G_1
 + \hat q G_2 + \hat p' G_3 + \hat p' \hat q G_4],
\end{equation}
where right hand side is nothing else but the decomposition of the
pseudoscalar bound-state amplitude into the independent matrix
structures. $G_j = G_j(p^2, p'^2, q^2)$ with $j=1,2,3,4$ and $i$
is the flavor index. Absolutely in the same way can be evaluated
the BS equation on account of the second skeleton diagram in Fig.
7 and so on. The actual evaluation of the BS equation for the pion
bound-state amplitude deserves a separate investigation as well as
the investigation of the Goldstone sector in QCD.

 \vfill

\eject

\end{document}